\documentclass[oneside]{article}

\usepackage[utf8]{inputenc}
\usepackage[a4paper, margin=4cm]{geometry}
\usepackage{parskip}
\usepackage{comment}
\usepackage{amsmath} 
\usepackage{amssymb}
\usepackage{enumerate}
\usepackage[english]{babel}
\usepackage{tipa}
\usepackage{amsmath}
\usepackage{wasysym}
\usepackage{makecell}
\usepackage{listings}
\usepackage{url}
\usepackage{mathtools}
\usepackage{amsthm}
\usepackage[colorlinks=true,
            linkcolor=blue,
            citecolor=blue,
            urlcolor=blue]{hyperref}
\usepackage{cleveref}
\usepackage{multirow}
\usepackage{subfig,graphicx}
\usepackage[toc,page,title]{appendix}
\PassOptionsToPackage{table}{xcolor}
\usepackage{colortbl}
\usepackage{booktabs}
\usepackage[table]{xcolor}
\usepackage{array}
\usepackage{caption}
\usepackage[export]{adjustbox}
\usepackage[numbers,sort&compress]{natbib}
\usepackage{float}

\floatstyle{plaintop}
\restylefloat{table}
\numberwithin{equation}{section}

\DeclareMathOperator{\sinc}{sinc}
\DeclareMathOperator*{\argmax}{arg\,max}

\begin{document}

\title{Vegetation pattern formation induced by local growth outpacing susceptibility to non-local competition
\author{
Jelle van der Voort\thanks{Mathematical Institute, Leiden University, 2300 RA, Leiden, Netherlands} 
\quad 
Ricardo Martinez-Garcia\thanks{\parbox[t]{0.85\textwidth}{Center for Advanced Systems Understanding,(CASUS) – Helmholtz-Zentrum Dresden-Rossendorf (HZDR), Untermarkt 20, Görlitz, 02826, Germany}}\space\thanks{\parbox[t]{0.85\textwidth}{ICTP South American Institute for Fundamental Research \& Instituto de Física Teórica, Universidade Estadual Paulista- UNESP, R. Dr. Bento Teobaldo Ferraz, 271- 2- Várzea da Barra Funda, São Paulo- SP, 01140-070, Brazil}} 
\quad 
Arjen Doelman$^*$
}
}

\maketitle

\begin{abstract}
\noindent In this work, we present and analyze a general framework for vegetation dynamics in arid and semi-arid ecosystems in which non-local interactions are purely competitive. The generality of the formulation enables a systematic search for ecological mechanisms that may lead to self-organized patterns. We identify two distinct mechanisms generating Turing instabilities across a broad class of models. The first mechanism arises from intensified competition in the areas between vegetated patches due to the cumulative pressure from their surroundings, and is well-documented in the literature. The second mechanism is novel and occurs when local growth outpaces competitive susceptibility near the uniform equilibrium. The analytical findings are complemented by numerical simulations of two benchmark models, both exhibiting a supercritical Turing bifurcation that leads to the formation of stable and robust vegetation patterns.
\end{abstract}

\vspace*{\fill}

Author for correspondence: Jelle van der Voort\\
Email: \textit{j.van.der.voort@math.leidenuniv.nl} 

\newpage

\section{Introduction}

The occurrence of regular vegetation patterns in arid and semi-arid ecosystems is a well-documented and widespread phenomenon \citep{Rietkerk2008,Borgogno2009,Meron2019,Gandhi2019,Clerc2021,Kastner2024}. These patterns are commonly attributed to self-organization among plants, driven by scale-dependent feedback between vegetation and resource availability \citep{Rietkerk2008}. Over the past decades, theoretical modelling has been remarkably successful in reproducing such patterning, thereby providing valuable insights into their underlying ecological and physical processes \citep{Borgogno2009,Meron2019,Landge2020,Martinez2023,Hou2025}. A widely studied class of models is based on reaction-diffusion systems, typically consisting of at least one vegetation and one water component \citep{Klausmeier1999,Rietkerk2002,Gilad2007,Marasco2014}. In such systems, vegetation growth depends on the local availability of water and the spatial redistribution of water occurs on a much faster timescale than vegetation growth and spread, enabling the scale-dependent feedback that is necessary for spatial instabilities to emerge. 

The separation of timescales between water and vegetation dynamics has led to the development of models in which the effect of water availability on vegetation growth is not represented explicitly \citep{Fuentes2004,Clerc2005,Martinez2013A,Segal2013,Martinez2013,Martinez2014,Tanzy2015,Simoy2023,Tega2022,Tlidi2020}. Instead, it is encoded in a kernel interaction function that accounts for how neighbouring vegetation impacts the growth of plants at a focal position \citep{Borgogno2009,Martinez2023}. These non-local models recover the same results predicted by local water-vegetation models, and indeed both can be reduced to a ``universal'' equation near the onset of the instability \citep{PintoRamos2025}. The rich phenomenology exhibited by models with non-local plant interactions motivated the use of similar non-local terms to describe other processes, including the impact of fire \citep{Tega2022,Patterson2024,Shen2025}, herbivory \citep{Siero2018}, seed dispersal \citep{Pueyo2008,Eigentler2018,Bennett2019,Tlidi2020,Bennett2023,Patterson2024,Shen2025} and delayed water absorption \citep{Liang2024,Xue2025}. A recent comprehensive review emphasizes the importance of non-local terms for improving the accuracy of mathematical models of physical and biological systems \citep{Pal2025}.

Some models explicitly distinguish between non-local facilitative and competitive effects \citep{Lefever1997,Tega2022,Tlidi2020}. For example, in the case of woody plants, facilitation may occur within the spatial extent of the canopy, where shading reduces evaporation and enhances water availability. Competition, on the other hand, is mediated mainly by water uptake via lateral root systems, which can extend far beyond the canopy \citep{Schenk2002,Barbier2008,Sternberg2005}. For this family of models, non-local self-limiting competition alone may generate spatial patterns, even when facilitation is either absent or restricted to the local limit \citep{Martinez2013,Martinez2014}. Several studies have demonstrated that such models produce patterning across a wide range of parameter values and kernel functions \citep{Britton1989,Fuentes2004,Clerc2005, Martinez2013A,Segal2013,Martinez2013,Martinez2014,Tanzy2015,Simoy2023}. A key mathematical requirement to obtain patterns in most of these models is that the Fourier transform of the competition kernel attains negative values. This condition is typically satisfied by kernels with a box-like shape, for which the competitive intensity remains relatively high and uniform over a finite spatial range before declining sharply to zero at greater distances \citep{Leimar2008,Pigolotti2010,Barabas2012,Leimar2013}. Ecologically, this requirement translates into the emergence of exclusion zones, bare soil regions surrounding vegetated patches where the establishment of new plants is suppressed \citep{Pigolotti2007,Pigolotti2010,Martinez2013,Martinez2013A}.

Vegetation patterning due to exclusion areas in models driven by non-local competition is well-documented in the literature. Nonetheless, a systemic analysis of whether alternative mechanisms within such models can also give rise to pattern formation is lacking. To address this gap, we develop a general modelling framework and analyze it to identify additional competition-induced mechanisms capable of generating patterns through a Turing bifurcation. Our objective is not only to establish the conditions under which instabilities can occur but also to provide a clear ecological interpretation. The use of a general modelling structure with unspecified functional forms is further motivated by recent findings indicating that simplistic biomass-density growth assumptions, such as the widespread use of logistic growth, may fail to capture essential ecological outcomes, including species coexistence \citep{Sauers2022,Spaak2023,Hatton2024}.

The remainder of this paper is structured as follows. In Section \ref{section: model}, we introduce the general modelling framework, which includes non-local self-limiting competition, general growth functions and biomass spread. We perform linear stability analysis in Section \ref{section: analysis}, and identify and analyze the distinct mechanisms that give rise to a Turing instability. Finally, in Section \ref{section: numerics}, we illustrate and extend the theoretical results through numerical simulations of two benchmark models.

\section{A general single-component model}
\label{section: model}

\subsection{Model formulation}
\label{section: model formulation}
We propose the following general modelling framework to describe vegetation dynamics across space, incorporating local growth, non-local self-limiting competition for resources and biomass spread:
\begin{align}
\label{model: general model}
    \frac{\partial u}{\partial t} = g(u) - s(u) \int \phi(x'-x)\, c(u(x',t))\, dx' + D \Delta u.
\end{align}
Here, $u(x,t)$ denotes the biomass density of a given vegetation type at location $x$ at time $t$. The function $g(u)$ represents all demographic and physiological processes that operate locally in space. Depending on the ecological context, $g(u)$ may take various forms, for example, linear, logistic, sublinear \citep{Hatton2024}, or any non-standard empirically fitted function. It is controlled by traits governing growth, self-limitation, and survival, such as maximum growth rate, photosynthetic capacity, nutrient-use efficiency, and stress tolerance.

Non-local competition is included via the integral term, which accounts for the cumulative competitive effects exerted by surrounding vegetation. The functions responsible for this spatial coupling, $s(u)$ and $c(u)$, can be interpreted in terms of plant traits using the standard decomposition of competition into effect and response components \citep{Goldberg1991,Violle2007}. 

The competitive pressure $c(u)\geq 0$ describes the competitive impact exerted by biomass on its surrounding. It is associated with traits that determine the strength and spatial extent of resource uptake and interference. These include root-associated traits such as its lateral spread, and resource mining or allelopathic potential \citep{Cabal2020,Britto2021,Cabal2024}. We impose $c'(u) > 0$, as the competitive pressure should intensify with biomass density, reflecting the extensive root systems with increased resource uptake capabilities of mature vegetation \citep{Milton1995,Schwinning1998,Schenk2002,Sea2012}. Although $c(u)$ is often assumed to be linear for analytical convenience, this assumption oversimplifies the inherently non-linear nature of competitive interactions.

Finally, the susceptibility function $s(u) \geq 0$ represents the sensitivity of local biomass to neighbor-induced resource depletion. Empirically, this function is related to tolerance and allocation traits, such as shade tolerance and biomass allocation patterns. For example, a higher root-to-shoot ratio or deeper rooting depth increases access to belowground resources under crowding and would typically reduce competitive susceptibility, corresponding to a lower $s(u)$. Typically, $s(u)$ should increase more rapidly for small $u$, reflecting the greater vulnerability to resource competition of early life-stage plants with shallow root systems. For larger $u$, slowing down or saturation is plausible, corresponding to established vegetation that is more resilient due to a robust and extensive root system \citep{Milton1995,Schwinning1998,Schenk2002,Sea2012}.

A variety of functional forms for $g(u)$, $s(u)$ and $c(u)$ have been studied in the literature \citep{Clerc2005,Tanzy2015,Li2020,Kavallaris2023,Tlidi2024}, motivating our choice to adopt a general formulation. We impose $g(0)=s(0)=c(0)=0$ to ensure that all processes cease in the absence of vegetation. This assumption also preserves non-negativity of solutions for ecologically meaningful initial conditions. Moreover, both $s(u)$ and $c(u)$ should equal zero only at $u=0$, as the effects of competition are inherently present when vegetation exists.

The kernel function $\phi(x)$ governs the spatial extent of non-local competition and will be discussed in detail in Section \ref{section: the kernel function phi}. 

Spatial biomass spread is modelled via diffusion, with diffusion coefficient $D>0$. This formulation provides a simplified representation of biomass spread through processes such as seed dispersal or clonal growth, as commonly employed in spatial ecological models \citep{Klausmeier1999,Rietkerk2002,Gilad2007,Marasco2014}. Mathematically, it can be recovered from a more realistic non-local dispersal kernel by truncating its power series expansion at second order \citep{Eigentler2018,Surendran2025}. We adopt the diffusion approximation both for analytical tractability and to demonstrate that spatial pattern formation does not require any non-local facilitative interactions \citep{Martinez2013}.

A particular case that can be recovered within this general framework is the non-local Fisher-KPP model (see also Section \ref{section: Fisher-KPP model set-up}) and its derivatives \citep{Bian2017,Li2020,Kavallaris2023}, which have been extensively used in spatial ecology \citep{Fuentes2004,Hernandez2004,Leimar2013,Paulau2014,Genieys2006,Silvano2025}. Other non-local models of population dynamics available in the literature \citep{Britton1989,Clerc2005} also fit naturally within our general formulation.

\subsection{\texorpdfstring{The kernel function $\phi(x)$}{The kernel function phi(x)}}
\label{section: the kernel function phi}

The kernel function $\phi(x)$ characterizes how the competitive pressure of vegetation is distributed across space. Since our focus is on the identification of competition-driven mechanisms that generate Turing patterns, we exclude any form of net non-local facilitation. Mathematically, this corresponds to assuming the kernel to be non-negative: $\phi(x) \geq 0$. 

Given that plants have a limited spatial range of influence, we require $\lim_{x \rightarrow \pm\infty} \phi(x) = 0$ to ensure that competition becomes negligible at large distances. Furthermore, we focus exclusively on kernels with compact support on the interval $[-\ell,\ell]$, where $\ell$ denotes the interaction range. This assumption excludes kernels with infinite tails, such as the Gaussian and Laplacian kernel, which are more suitable to describe processes like seed dispersal, where long distance transport, although rare, does occur \citep{Nathan2012,Eigentler2018,Pueyo2008,Bennett2019}. In contrast, non-local (resource) competition is inherently more localized as root systems are spatially limited and lateral water movement typically occurs over short distances \citep{Schenk2002}. Therefore, it is ecologically reasonable to assume that competition vanishes beyond a certain critical distance. 

To reflect the decay of competitive influence farther from the focal point, we further assume $\phi(x)$ to be non-increasing on $[0, \ell]$. Additionally, we assume a spatially homogeneous environment (for simplicity), resulting in isotropic competitive effects. This is ensured by requiring the kernel to be symmetric: $\phi(x)=\phi(-x)$. As a last condition, we assume the kernel to be normalized: $\int \phi(y) dy = 1$. This allows for a direct correspondence between the spatially explicit model and its mean field (spatially homogeneous) counterpart. This normalization can be imposed on any ecologically relevant kernel by scaling with an appropriate constant.

\begin{figure}[t!]
    \centering
    \begin{minipage}{\textwidth}
        \centering
        \includegraphics[width=\textwidth]{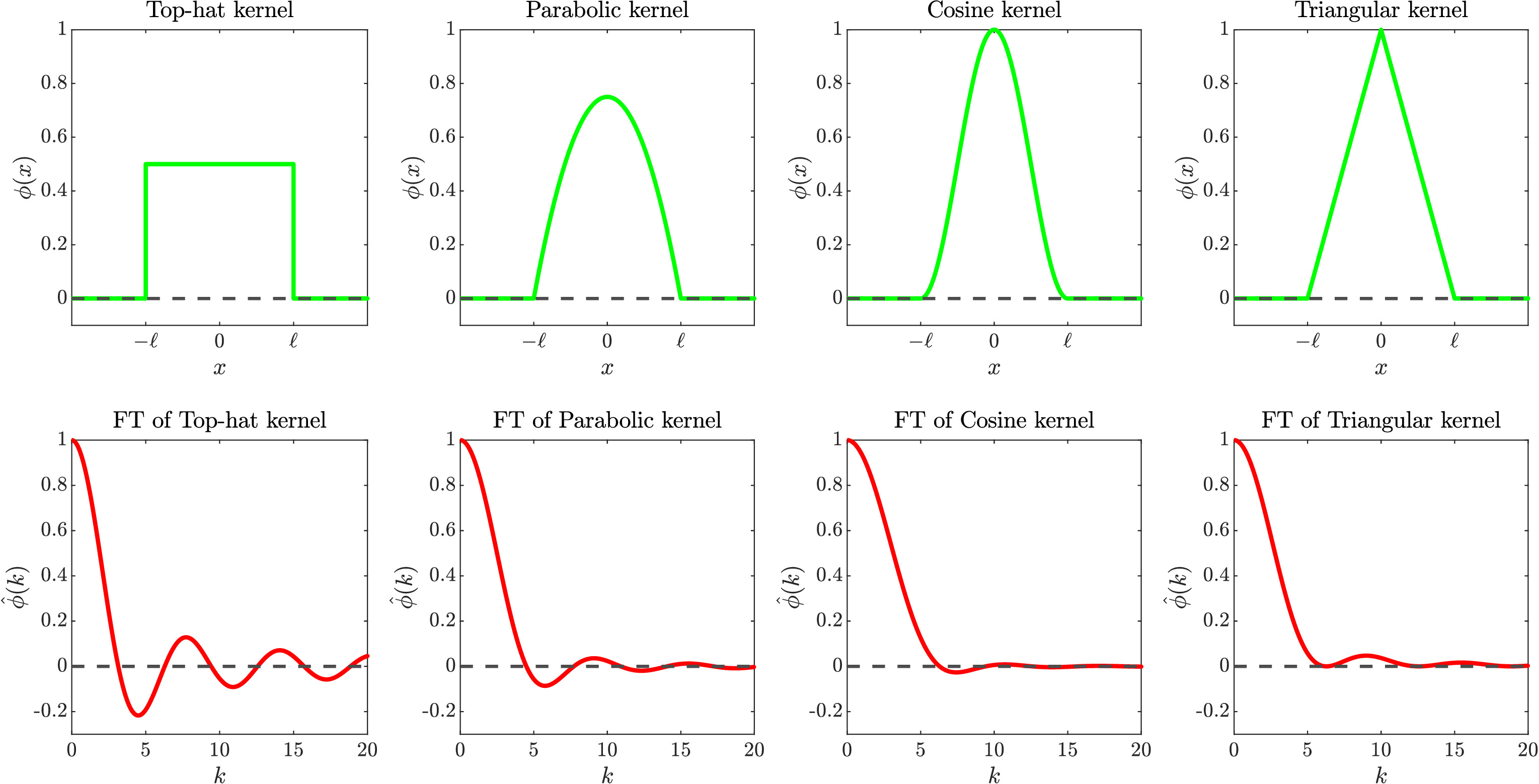}
    \end{minipage}%
\caption{Plots of the top-hat, parabolic, cosine and triangular kernel (top row) and their Fourier transforms (bottom row).}
\label{figure: representative kernels}
\end{figure}

A variety of kernel functions satisfying these criteria have been used in earlier studies \citep{Fuentes2003,Pigolotti2007,Clerc2005,Segal2013}, with the top-hat kernel being a particularly common choice due to its simplicity \citep{Hernandez2004,Martinez2013A,Martinez2013,Andreguetto2021,Piva2021,Tega2022,Simoy2023,Li2025}. In our analysis, we maintain generality by leaving $\phi(x)$ unspecified, subject to the above assumptions. For some analytical comparisons and in the numerical experiments, we focus on the four representative examples shown in the top row of Figure \ref{figure: representative kernels}: the top-hat, the parabolic, the cosine and the triangular kernel, each parametrized by a single parameter: the interaction range $\ell$.

The top-hat kernel is the simplest, with the competition intensity constant within the whole interaction range. The other three kernels do account for competition strength decaying toward the edges, but assume different spatial profiles. The parabolic kernel emphasizes strong competition near the center, with a progressively increasing rate of decline toward the edges. The cosine and triangular kernel are more sharply peaked at the center, resembling a narrow pulse. Compared to the triangular kernel, which has a constant rate of decline, the cosine kernel has more mass near the origin and less mass around its tail. Full expressions for, and derivations of these kernels are provided in Appendix \ref{appendix: kernel functions}.

The specific choice of kernel can be informed by available empirical data \citep{Cabal2020}. For example, if a plant's root system is distributed relatively evenly around the stem, a box-like kernel such as the top-hat kernel may be appropriate. On the other hand, if the roots are concentrated near the crown with a few extensions reaching farther out, the cosine kernel may better capture the natural root distribution.

To facilitate the forthcoming analysis, we also introduce the Fourier transform of the kernel function $\phi(x)$:
\begin{align}
\label{definition: FT of phi}
    \hat{\phi}(k) &= \int_{-\infty}^{\infty} \phi(x)\, e^{ikx} \, dx 
    \overset{\substack{\text{compact} \\ \text{support}}}{=} 
    \int_{-\ell}^{\ell} \phi(x)\, e^{ikx} \, dx 
    \overset{\substack{\text{symmetry}}}{=} 
    2\int_{0}^{\ell} \phi(x)\, \cos(kx) \, dx .
\end{align}
The properties of $\hat{\phi}(k)$, particularly whether or not a sign change occurs, play a central role in the upcoming analysis. As shown in the bottom row of Figure \ref{figure: representative kernels}, the Fourier transforms of the top-hat, the parabolic and the cosine kernel take dips below zero, while the Fourier transform of the triangular kernel remains non-negative for all $k$ (see Appendix \ref{appendix: kernel functions} for more details). Although no simple criterion exists that is both necessary and sufficient for the positivity of Fourier transforms, some conditions addressing the issue have been established \citep{Tuck2006,Giraud2014}. A frequent cause of a sign changing Fourier transform is a kernel that decays slowly near the center but exhibits sharp drops further away \citep{Leimar2008,Pigolotti2010,Barabas2012,Leimar2013}. 

Moreover, $\hat{\phi}(0)=1$ due to the normalization condition and the symmetry of $\phi(x)$ implies that $\hat{\phi}(k)$ is real-valued and symmetric, as follows from (\ref{definition: FT of phi}). Finally, the Riemann-Lebesgue Lemma yields $\lim_{k \rightarrow \pm \infty} \hat{\phi}(k) = 0$ \citep{Serov2017}.

\section{Near the onset of Turing patterns}
\label{section: analysis}

\subsection{Linear stability analysis}
We derive general conditions under which the model (\ref{model: general model}) admits an ecologically relevant uniform steady state that can undergo a Turing instability. As a first step, we consider the existence of spatially homogeneous equilibria $\bar{u}>0$, representing uniformly vegetated states. These equilibria can be obtained by substituting $\bar{u}(x,t)=\bar{u}$ into (\ref{model: general model}), which yields the (uniform) steady state equation,
\begin{align}
\label{equation: steady state}
    g(\bar{u})=s(\bar{u}) c(\bar{u}).
\end{align}
Naturally, arbitrary choices for the model functions $g(u)$, $s(u)$ and $c(u)$ do not guarantee the existence of a positive steady state. Nevertheless, in many ecologically plausible scenarios, local growth dominates at low and/or intermediate biomass densities, while competition for resources becomes limiting at higher densities. This typically ensures the existence of a positive intersection point between the functions $g(u)$ and $s(u) c(u)$. Therefore, we simply proceed under the assumption that a positive equilibrium $\bar{u} > 0$ exists for the chosen model functions. Furthermore, the positivity of $s(\bar{u})$ and $c(\bar{u})$ implies $g(\bar{u})> 0$ as well.
 
The stability of $\bar{u}$ against both homogeneous and heterogeneous perturbations can be analyzed by employing the Fourier ansatz,
\begin{align}
  u(x,t) = \bar{u} + \varepsilon e^{\omega t+i k x} +\rm{c.c.},
\end{align}
where $k$ is the wave number, $\varepsilon>0$ should be sufficiently small and $\rm{c.c.}$ denotes the complex conjugate. The dispersion relation $\omega=\omega(k)$ represents the growth rate of the perturbation mode corresponding to the wave number $k$. Substituting this expression into (\ref{model: general model}) and linearizing the functions $g(u)$, $s(u)$ and $c(u)$ around $\bar{u}$ yields,
\begin{align}
\omega \cdot \varepsilon e^{\omega t + i k x} &=
g(\bar{u}) + g'(\bar{u}) \cdot \varepsilon e^{\omega t + i k x} - \left(s(\bar{u}) + s'(\bar{u}) \cdot \varepsilon e^{\omega t + i k x}\right) \nonumber \\
&\quad \cdot \left(c(\bar{u}) + c'(\bar{u}) \int \phi(x' - x) \, e^{i k x'}\, dx' \cdot \varepsilon e^{\omega t} \right) - D k^2 \cdot \varepsilon e^{\omega t + i k x} + \mathcal{O}(\varepsilon^2).
\end{align}
The integral can be rewritten as follows,
\begin{align}
    \int \phi(x' - x) e^{i k x'}\, dx' = e^{ikx} \int \phi(x' - x) e^{i k (x'-x)}\, dx' = e^{ikx} \int \phi(y) e^{i k y}\, dy = e^{ikx} \hat{\phi}(k).
\end{align}
Invoking the steady state equation (\ref{equation: steady state}), we find that the leading order equation is of order $\varepsilon$, yielding an expression for the dispersion relation,
\begin{align}
     \omega(k) &= g'(\bar{u}) - s'(\bar{u})c(\bar{u}) - s(\bar{u})c'(\bar{u})\hat{\phi}(k)-Dk^2.
\end{align}
We can rewrite the first two terms by employing the steady state equation (\ref{equation: steady state}) and the fact that $s(\bar{u})>0$,
\begin{align}
    g'(\bar{u}) - s'(\bar{u})c(\bar{u}) 
    &= g'(\bar{u}) -s'(\bar{u})\frac{g(\bar{u})}{s(\bar{u})} = s(\bar{u}) \left.\frac{d}{du}\left(\frac{g(u)}{s(u)}\right)\right\rvert_{u=\bar{u}},
\end{align}
so that,
\begin{align}
\label{equation: dispersion relation}
     \omega(k) &= s(\bar{u})\left(\left.\frac{d}{du}\left(\frac{g(u)}{s(u)}\right)\right\rvert_{u=\bar{u}}-c'(\bar{u})\hat{\phi}(k)\right)-Dk^2.
\end{align}
For the equilibrium $\bar{u}$ to be Turing unstable, it must be temporally stable (i.e., stable against perturbations that are homogeneous in space) but spatially unstable (i.e., unstable against perturbations that are heterogeneous in space). In other words, the growth rate $\omega(k)$ must be negative at $k=0$, yet positive for at least some finite range of wavenumbers. Moreover, for the problem to be well-posed (i.e., for the system to suppress the appearance of exponentially growing perturbations with arbitrarily small wavelengths), it is required that $\omega(k) < 0$ as $k \rightarrow \infty$. Since $\hat{\phi}(k) \rightarrow 0$ as $k \rightarrow \infty$, the diffusive component $-Dk^2$ ensures that this condition holds without additional constraints. This effect confirms the stabilizing role of biomass spread, even when the diffusion rate $D$ is arbitrarily small yet strictly positive. This behaviour contrasts with the destabilizing role that diffusion can play in two-component reaction-diffusion systems \citep{Doelman2019}.

Assessing temporal stability, we insert $k=0$ into (\ref{equation: dispersion relation}), which yields the condition,
\begin{align}
\label{equation: temporal stability}
    \left.\frac{d}{du}\left(\frac{g(u)}{s(u)}\right)\right\rvert_{u=\bar{u}} < c'(\bar{u}).
\end{align}
To trigger a spatial instability, non-local self-limiting competition must overcome the stabilizing effect of diffusion, i.e., it must outweigh the negative contribution of diffusion to the dispersion relation. In mathematical terms, the following necessary condition for a spatial instability follows from (\ref{equation: dispersion relation}),
\begin{align}
\label{equation: necessary condition spatial instability}
    c'(\bar{u})\inf_k{\hat{\phi}(k)} &< \left.\frac{d}{du}\left(\frac{g(u)}{s(u)}\right)\right\rvert_{u=\bar{u}}.
\end{align}
This condition becomes sufficient when the diffusion coefficient $D$ is sufficiently small,
\begin{align}
\label{equation: Dmax}
    D< D_\text{max} \coloneqq s(\bar{u}) \cdot\max_{k>0} \left( \frac{\left.\frac{d}{du}\left(\frac{g(u)}{s(u)}\right)\right\rvert_{u=\bar{u}}-c'(\bar{u})\hat{\phi}(k)}{k^2}\right),
\end{align}
where $D_\text{max}$ is well-defined provided that the condition in (\ref{equation: necessary condition spatial instability}) is satisfied. This threshold also marks the location of the Turing bifurcation, with associated critical wave number $k=k_c$,
\begin{align}
\label{equation: kc}
    k_c \coloneqq \argmax_{k>0} \left(\frac{\left.\frac{d}{du}\left(\frac{g(u)}{s(u)}\right)\right\rvert_{u=\bar{u}}-c'(\bar{u})\hat{\phi}(k)}{k^2}\right).
\end{align}

\begin{figure}[t!]
    \centering
    \begin{minipage}[b]{\textwidth}
        \centering
        \includegraphics[width=0.5\textwidth]{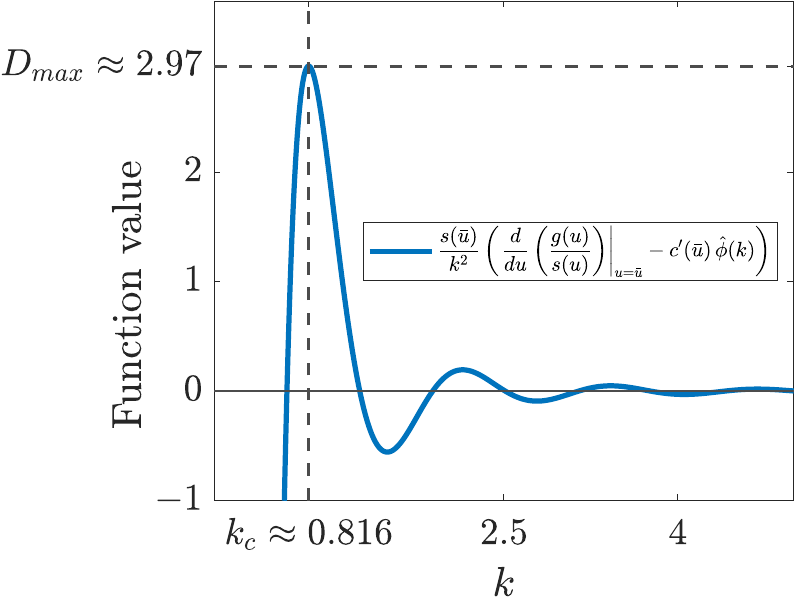}
    \end{minipage}%
\caption{$D_\text{max}$ and $k_c$ for the non-local Fisher-KPP model (see (\ref{model: non-local Fisher KPP})) with a top-hat kernel for $a = 10$ and $\ell = 5$.}
\label{figure: D_max}
\end{figure}

Figure \ref{figure: D_max} illustrates an example of the function in (\ref{equation: Dmax}) to be maximized, with $D_\text{max}$ corresponding to its maximum and $k_c$ to the wavenumber at which it is attained.

\subsection{Mechanisms for a Turing instability}

From the linear stability analysis it follows that a Turing instability is present in the model (\ref{model: general model}) if there exists a homogeneous equilibrium $\bar{u}>0$ satisfying
\begin{align}
\label{equation: Turing instability condition}
    \inf_{k>0} \hat{\phi}(k) < \frac{1}{c'(\bar{u})} \left.\frac{d}{du}\left(\frac{g(u)}{s(u)}\right)\right\rvert_{u=\bar{u}} < 1,
\end{align}
together with $D < D_\text{max}$. We observe that if both
\begin{align}
\inf_{k>0} \hat{\phi}(k) = 0 \quad \text{and} \quad \left.\frac{d}{du}\left(\frac{g(u)}{s(u)}\right)\right\rvert_{u=\bar{u}} \leq 0
\end{align}
hold, then the left inequality in (\ref{equation: Turing instability condition}) is not satisfied, excluding the possibility of a Turing bifurcation. Therefore, we focus on three distinct scenarios: 1) $\hat{\phi}(k)$ takes an excursion below zero (Section \ref{subsubsection: exclusion areas}), 2) $\frac{g(u)}{s(u)}$ is increasing at the homogeneous equilibrium $\bar{u}$ (Section \ref{subsubsection: GOS}), 3) both conditions occur simultaneously (Section \ref{subsubsection: both}). Below, we work out the mathematical details for these cases and discuss their ecological interpretation.

\subsubsection{Stronger competitive pressure in between high biomass areas}
\label{subsubsection: exclusion areas}

First, we investigate whether a Turing instability can arise under the constraint 
\begin{align}
    \left.\frac{d}{du}\left(\frac{g(u)}{s(u)}\right)\right\rvert_{u=\bar{u}}\leq0,
\end{align}
which yields that the the second inequality in (\ref{equation: Turing instability condition}) holds without any additional requirements. Therefore, we are interested in whether the following inequality can be satisfied,
\begin{align}
\label{condition: exclusion zone}
    \inf_{k>0} \hat{\phi}(k) < \frac{1}{c'(\bar{u})} \left.\frac{d}{du}\left(\frac{g(u)}{s(u)}\right)\right\rvert_{u=\bar{u}}.
\end{align}
Since we assumed that $c'(u) > 0$ (Section \ref{section: model formulation}), it follows that $\inf_{k>0} \hat{\phi}(k)<0$ is a necessary condition. Furthermore, if the competitive impact increases steeply near the equilibrium, the right-hand-side approaches zero, thereby increasing the likelihood of the inequality to be satisfied. The same conclusion applies when the ratio of local growth to competitive susceptibility is only weakly decreasing close to the homogeneous equilibrium. Generally, both effects can be tuned through the model parameters, allowing the right-hand-side to approach zero from below. 

For instance, considering logistic growth combined with linear functions for both the competitive susceptibility and impact, i.e.,
\begin{align}
    g(u) &= a u \left(1 - \frac{u}{K}\right),
    &\quad s(u) &= bu,
    &\quad c(u) &= r u,
\end{align}
the condition in (\ref{condition: exclusion zone}) becomes
\begin{align}
     \inf_{k>0} \hat{\phi}(k) < -\frac{a}{bKr}.   
\end{align}
By decreasing $a$ or increasing $b$, $K$ or $r$, the right-hand-side can be brought arbitrarily close to zero, so that the inequality can be satisfied whenever $\hat{\phi}(k)$ takes a dip below zero.

In summary, the sign of $\inf_{k>0} \hat{\phi}(k)$ generally determines the theoretical possibility for a Turing bifurcation. From an ecological perspective, it can be verified whether the instability occurs within a biologically relevant region of the parameter space by employing a parametrization procedure based on empirical data.

\begin{figure}[b!]
    \centering
    \begin{minipage}[b]{0.49\textwidth}
        \centering
        \includegraphics[width=\textwidth]{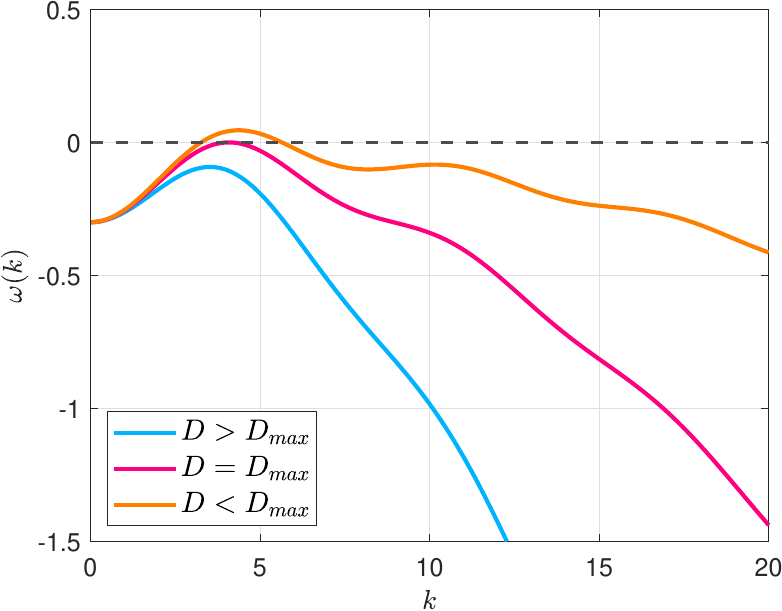}
    \end{minipage}%
    \hspace{0.15cm}
    \begin{minipage}[b]{0.49\textwidth}
        \centering
        \includegraphics[width=\textwidth]{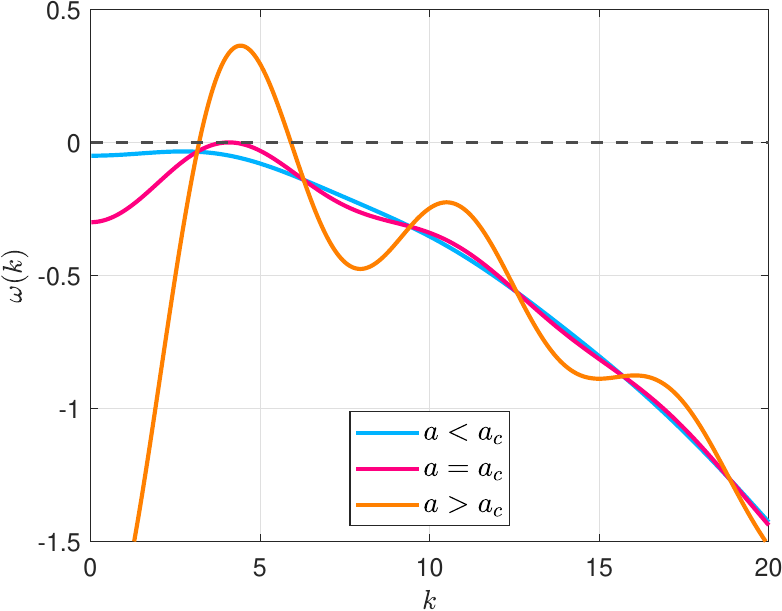}
    \end{minipage}%
\caption{The dispersion relation $\omega(k)$ for the non-local Fisher-KPP model with a top-hat kernel ($\ell = 1$), local growth rate $a = 0.3$ and diffusion coefficient $D = 0.0035628$. A Turing bifurcation occurs when $\omega(k)$ crosses zero, either as $D$ is decreased (left panel) or as $a$ is increased (right panel).}
\label{figure: dispersion relation}
\end{figure}

Continuing under the assumption that $\hat{\phi}(k)$ changes sign, we illustrate the typical behaviour of the dispersion relation (\ref{equation: dispersion relation}) when varying a bifurcation parameter. We assume a general shape of $\hat{\phi}(k)$ similar to the Fourier transforms shown in the bottom row of Figure \ref{figure: representative kernels}: oscillatory, with countably many excursions below zero, and amplitudes that decay as $|k|$ increases. At the Turing bifurcation point (i.e., for the critical wave number $k=k_c$), we require $\omega(k_c) = \omega'(k_c)=0$ \citep{Doelman2019}. This implies that the dispersion relation evaluated at the bifurcation point takes the qualitative form as shown in magenta in Figure \ref{figure: dispersion relation}, where the curve touches the $k$-axis at its first and highest maximum.

The contribution of diffusion to the dispersion relation, i.e., $-Dk^2$, is strictly negative, providing the aforementioned stabilizing effect. In particular, it ensures that $\omega(k)$ decays quadratically for large $k$ (as $\hat{\phi}(k) \rightarrow 0$ when $k \rightarrow \infty$). Reducing the diffusion coefficient $D$ delays the onset of this quadratic decay to larger $k$, potentially allowing $\omega(k)$ to cross the $k$-axis and triggering a Turing bifurcation, as illustrated in the left panel of Figure \ref{figure: dispersion relation}.

Any of the intrinsic model parameters, captured within the functions $g(u)$, $s(u)$ and $c(u)$, can also serve as a bifurcation parameter. As an example, for the non-local Fisher-KPP model (see (\ref{model: non-local Fisher KPP})), the dispersion relation reduces towards,
\begin{align}
    \omega(k) = -a \frac{\hat{\phi}(k)}{k^2}-Dk^2,
\end{align}
with $a$ depicting the local growth rate. Larger values of $a$ increase the maximum of the non-diffusive part of the dispersion relation. As $a$ exceeds the critical threshold $a_c$, the peak of the first term becomes greater than the stabilizing effect of diffusion, thereby generating a Turing instability, as shown in the right panel of Figure \ref{figure: dispersion relation}.

The ecological interpretation of the vegetation patterning induced by the mathematical condition $\inf_{k>0} \hat{\phi}(k)<0$ is commonly linked to the emergence of exclusion zones \citep{Pigolotti2007,Pigolotti2010,Martinez2013,Martinez2013A}. Within these bare soil regions, the establishment of new individuals is highly unlikely as seedlings would need to compete with vegetation from two surrounding patches, while plants within each patch face competition only from their immediate neighbours. This mechanism is especially pronounced for box-like kernels with a relatively flat center, which amplify the cumulative impact of competition in between patches beyond the competitive pressure exerted within a single patch.

However, the full development of such exclusion areas is not always expected to occur immediately at the onset of a Turing instability. A decisive factor is whether the Turing bifurcation is subcritical or supercritical \citep{Cross1993,vanderstelt2013,Doelman2019}. A subcritical Turing bifurcation triggers the formation of unstable, transient patterns with increasing amplitude, which may drive the system towards a variety of possible outcomes, such as an alternative stable homogeneous state or a far-from-homogeneous-equilibrium pattern \citep{Krause2024,Voort2025}. The latter case can lead to the emergence of exclusion zones immediately beyond the bifurcation point, if the patterned transient drives the system towards a strongly localized patterned state characterized by regions of near-zero biomass, as observed in the models studied in \citep{Pigolotti2007,Pigolotti2010,Martinez2013,Martinez2013A}.

On the other hand, a supercritical Turing bifurcation generates small-amplitude stable patterns in which regions of zero biomass do not occur: vegetation persists across the entire domain, exhibiting oscillatory variations in density. We explore the ecological interpretation of these type of patterns in our numerical experiments.

\subsubsection{Growth outpacing susceptibility to competition}
\label{subsubsection: GOS}
The second scenario we consider is the possibility for a Turing instability under the restriction $\inf_{k>0} \hat{\phi}(k) = 0$ (recall that $\hat{\phi}(k) \rightarrow 0$ as $k \rightarrow \infty$ so that $\inf_{k>0} \hat{\phi}(k) \leq 0$). Then, the condition (\ref{equation: Turing instability condition}) reduces to
\begin{align}
\label{equation: Turing instability by GOS}
    0 < \left.\frac{d}{du}\left(\frac{g(u)}{s(u)}\right)\right\rvert_{u=\bar{u}} < c'(\bar{u}),
\end{align}
for a homogeneous equilibrium $\bar{u}>0$ defined by
\begin{align}
    \frac{g(\bar{u})}{s(\bar{u})} = c(\bar{u}).
\end{align}
That is, at the intersection point $u=\bar{u}$ of the curves $\frac{g(u)}{s(u)}$ and $c(u)$, the function $\frac{g(u)}{s(u)}$ must be increasing (for a spatial instability), but with a slope smaller than that of $c(u)$ (for temporal stability). Several clear-cut statements on Turing patterning follow directly. If $\frac{g(u)}{s(u)}$ is decreasing for all $u$, a Turing instability is excluded. If $\frac{g(u)}{s(u)}$ is increasing for all $u$, then any temporally stable steady state undergoes a Turing bifurcation for $D$ sufficiently small. If $\frac{g(u)}{s(u)}$ is increasing for smaller $u$, but decreasing for larger $u$ (a typical scenario when local growth includes a carrying capacity), it is required that
\begin{align}
    \bar{u} < u_\text{max} \coloneqq \argmax_{u>0} \left(\frac{g(u)}{s(u)}\right),
\end{align}
for Turing pattern formation to be possible. 

\begin{figure}[b!]
    \centering
    \begin{minipage}[b]{\textwidth}
        \centering
        \includegraphics[width=0.5\textwidth]{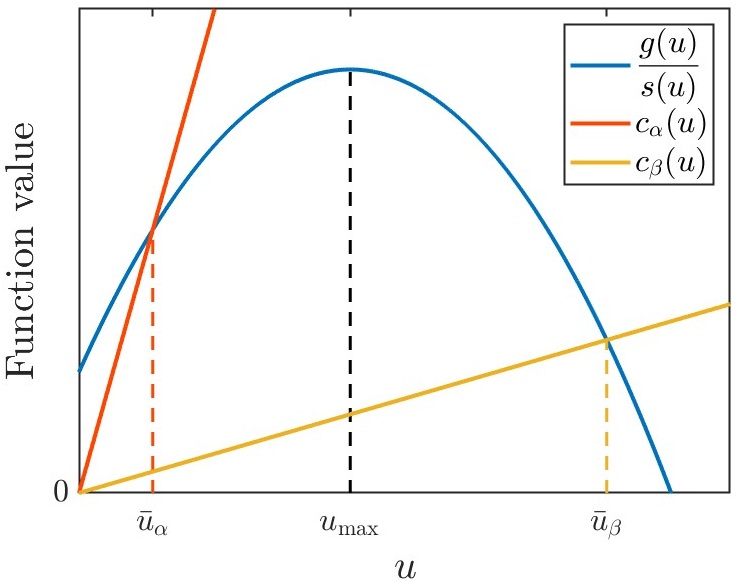}
    \end{minipage}%
\caption{The crossing of $\frac{g(u)}{s(u)}$ with $c_\alpha(\bar{u})$ gives rise to a Turing instability for sufficiently small $D$, whereas the crossing of $\frac{g(u)}{s(u)}$ with $c_\beta(\bar{u})$ does not.}
\label{figure: growth trumping susceptibility example}
\end{figure}

As an illustrative example, we consider logistic local growth for $g(u)$, a Holling type II saturating functional response for the competitive susceptibility $s(u)$, and a linear density-dependent competition intensity $c_r(u)$:
\begin{align}
    g(u) &= a u \left(1 - \frac{u}{K}\right),
    &\quad s(u) &= \frac{b u}{1 + c u},
    &\quad c_r(u) &= r u,
\end{align}

with $a,K,b,c,r >0$. In this setting, any value of the slope $r$ of the competitive intensity yields an intersection point between $\frac{g(u)}{s(u)}$ and $c_r(u)$ corresponding to a temporally stable equilibrium, since $\left.\frac{d}{du}\left(\frac{g(u)}{s(u)}\right)\right\rvert_{u=\bar{u}_r} < c_r'(\bar{u}_r)$ holds generically. However, a Turing instability arises only if $\bar{u}_r < u_\text{max}$, which requires the competitive intensity $c_r(u)$ to increase sufficiently fast, as illustrated in Figure \ref{figure: growth trumping susceptibility example}. We expect a similar requirement to apply for other ecologically relevant functional forms, because the decay of $g(u)$ at high $u$ otherwise violates the first inequality in (\ref{equation: Turing instability by GOS}). Moreover, the specific functional form of $c_r(u)$ may be of minor importance: under the assumptions $c(0)=0$ and $c'(u)>0$, parameters can typically be tuned such that the point of intersection occurs before $u_\text{max}$.

Ecologically, the first inequality in (\ref{equation: Turing instability by GOS}) can be interpreted as requiring that, in the vicinity of the homogeneous equilibrium, increases in biomass are accompanied by a stronger increase in local growth compared to the increase in competitive susceptibility. The inequality can also be written as,
\begin{align}
    \frac{g'(\bar{u})}{g(\bar{u})} > \frac{s'(\bar{u})}{s(\bar{u})},
\end{align}
which makes it explicit that the relative rate of increase of local growth must exceed the relative rate of increase of competitive susceptibility. Fulfillment of this condition opens a pathway for spatially heterogeneous departures from the uniform steady state.

Several ecological processes can contribute to this imbalance. One example arises from the life-stage dependent traits of individual plants: when a seedling has successfully established and entered an advanced growth stage, it develops an extensive and robust root system, at which point its potential for further growth can outweigh the susceptibility to competition from neighbouring individuals. Another mechanism involves local self-facilitation of aggregated vegetation. Densely vegetated patches can enhance soil water infiltration, thereby promoting their own growth while reducing sensitivity to competition for water from nearby plants \citep{Eldridge2000,Ludwig2004,Rietkerk2008}. Another example occurs in savannas, where adult trees located near juveniles can provide protection against fire \citep{Holdo2005}. Nevertheless, in many scenarios, as biomass continues to increase beyond a certain threshold where many mature individuals cluster together, the balance between local growth and susceptibility to intraspecific competition may shift, with competitive susceptibility eventually dominating growth due to limited resources.

The second inequality in (\ref{equation: Turing instability by GOS}) ensures that the cumulative competitive impact is strong enough to prevent an escape from the uniform equilibrium via homogeneous, (initially) exponential growth. Consequently, the only way to leave the steady state is via spatial pattern formation induced by a Turing bifurcation, which may lead to either a stable patterned state or convergence to another stable structure via a patterned transient \citep{Doelman2019,vanderstelt2013,Voort2025}.

Taken together, these conditions align with the LALI-principle of Local Activation and Lateral Inhibition \citep{Meinhardt2000}. Locally, vegetation growth outpaces susceptibility to competition in the relative sense, thereby activating the formation of a biomass bump. This local increase inhibits growth in its surrounding through non-local competition, generating a positive feedback loop that supports pattern formation. 

In many integro-differential models that include non-local self-limiting competition, the particular scenario of scale-dependent feedback where growth locally outpaces competitive susceptibility is absent. This is because both processes are typically represented by specific simplified functions (such as linear or logistic), making $\inf_{k>0} \hat{\phi}(k) < 0$ a necessary condition for pattern formation via the Turing principle \citep{Fuentes2004,Segal2013,Martinez2013,Martinez2014,Martinez2013A,Tega2022}. However, in some models, this restriction is relaxed, for example when incorporating an Allee effect in the growth function \citep{Clerc2005} or aggregation \citep{Britton1989}.

\subsubsection{Interplay of mechanisms}
\label{subsubsection: both}

The two mechanisms discussed above are not mutually exclusive: the first mechanism arises from the choice of the kernel function, while the second mechanism is determined by the choice of the functional forms for growth and competition. When both conditions are satisfied, the mechanisms may act in concert, potentially inducing a Turing bifurcation in parameter regimes where neither mechanism alone would suffice. The conditions under which each mechanism arises are summarized in Table \ref{table: conditions for mechanisms}.

\begin{table}[t!]
\centering
\caption{Scheme to determine whether a temporally stable uniform equilibrium $\bar{u}$ is spatially unstable (provided that $D$ is sufficiently small) and by which mechanism.}
\label{table: conditions for mechanisms}
\begin{tabular}{lcc}
\toprule
 \textbf{Conditions} & $\left.\frac{d}{du}\left(\frac{g(u)}{s(u)}\right)\right\rvert_{u=\bar{u}} \leq 0$ & $\left.\frac{d}{du}\left(\frac{g(u)}{s(u)}\right)\right\rvert_{u=\bar{u}} > 0$ \\
\midrule
$\hat{\phi} (k)\geq0$ for all $k$ & no spatial instability & \makecell{yes; growth outpacing \\ susceptibility}  \\
$\hat{\phi}(k)<0$ for some $k$ & \makecell{possibly, check (\ref{condition: exclusion zone}); \\ strong competition \\ in between vegetated patches} & \makecell{yes; both mechanisms \\ present} \\
\bottomrule
\end{tabular}
\end{table}

\section{Numerical simulations}
\label{section: numerics}

In this final section, we numerically simulate two representative models to examine the qualitative similarities and differences in the vegetation patterns generated by the two distinct instability mechanisms.

\subsection{Two benchmark models}

\subsubsection{The non-local Fisher-KPP model}
\label{section: Fisher-KPP model set-up}
The first model we consider is a non-local version of the Fisher-KPP model \citep{Fuentes2003}. The governing equation is given by,
\begin{align}
\label{model: non-local Fisher KPP}
    \dot{u} = au - bu \int \phi(x'-x)\, u(x',t)\, dx' + D \Delta u,
\end{align}
with parameters $a, b, D > 0$. This model represents a minimal example within our general framework, as the functions $g(u)$, $s(u)$ and $c(u)$ are all linear. It admits a spatially uniform vegetated steady state given by $\bar{u} = \frac{a}{b}$. Furthermore,
\begin{align}
    \left.\frac{d}{du}\left(\frac{g(u)}{s(u)}\right)\right\rvert_{u=\bar{u}} &= 0,
\end{align}
ensuring temporal stability via (\ref{equation: temporal stability}). Following Table \ref{table: conditions for mechanisms}, this indicates the possibility for a spatial instability only for competition kernels with a sign-changing Fourier transform. Condition (\ref{condition: exclusion zone}) reads
\begin{align}
    \inf_{k>0} \hat{\phi}(k) < 0,
\end{align}
indicating no additional parameter restriction for $\bar{u} = \frac{a}{b}$ to be Turing instable besides the requirement that $D$ should be smaller than the threshold value $D_\text{max}$ given by
\begin{align}
    D_\text{max} = -a \cdot \min_{k>0}\frac{\hat{\phi}(k)}{k^2}.
\end{align}

\subsubsection{The GOS model (Growth Outpacing Susceptibility)}

The second model represents a system in which the ratio of local growth to competitive susceptibility is continuously increasing with biomass. The dynamics are given by,
\begin{align}
\label{model: no facilitation}
    \dot{u} = au -\frac{bu}{1+cu} \int \phi(x'-x)\, u^2(x',t)\, dx' + D \Delta u,
\end{align}
with parameters $a, b, c, D > 0$. As in the non-local Fisher-KPP model, local growth is modelled by a linear term. In contrast, the competitive is sublinear, represented by a Holling type II functional response. This assumption implies that the sensitivity to competition increases more slowly as biomass accumulates. Furthermore, it ensures that the ratio of local growth to competitive susceptibility is increasing everywhere, for any parameter choice:
\begin{align}
\label{GOS model GOS condition}
\frac{d}{du}\left(\frac{g(u)}{s(u)}\right) = \frac{ac}{b} >0.
\end{align}

The competitive impact is quadratic in $u$, ensuring that early life-stage vegetation exerts little effect on its surroundings, while impact increases rapidly with maturity due to an extensive, well-established root system. 

The uniformly vegetated equilibrium is given by,
\begin{align}
    \bar{u} = \frac{ac}{2b}\left(1+\sqrt{1+\frac{4b}{ac^2}}\right),
\end{align}
which is positive for any choice of the model parameters. The condition for temporal stability in (\ref{equation: temporal stability}) evaluates to,
\begin{align}
    \left(1+\sqrt{1+\frac{4b}{ac^2}}\right)^{-1}<1,
\end{align}
which is clearly satisfied. Furthermore, the inequality in (\ref{GOS model GOS condition}) holds for all $u$, in particular for $u = \bar{u}$. Accordingly, following the scheme in Table \ref{table: conditions for mechanisms}, the equilibrium $\bar{u}$ is Turing unstable for any particular choice of kernel, whenever $D < D_\text{max}$, with
\begin{align}
    D_\text{max} = 2a \cdot \max_{k>0} \frac{\left(1+\sqrt{1+\frac{4b}{ac^2}}\right)^{-1}-\hat{\phi}(k)}{k^2}.
\end{align}

\subsubsection{Kernel functions inducing a Turing instability}

\begin{table}[b!]
\centering
\caption{The specific choice of competition kernel determines whether or not the non-local Fisher-KPP model and the GOS model admit a positive homogeneous equilibrium that is Turing unstable, and by which mechanism this instability arises, provided that $D < D_\text{max}$ holds.}
\label{table: conditions for specific models and kernels}
\begin{tabular}{lcc}
\toprule
 \textbf{Kernel/Model} & Fisher-KPP & GOS \\
\midrule
triangular & no Turing instability  & \makecell{yes; growth outpacing \\ susceptibility}  \\
top-hat, parabolic, cosine & \makecell{yes; strong competition \\ in between vegetated patches} & \makecell{yes; both mechanisms \\ present} \\
\bottomrule
\end{tabular}
\end{table}

In Table \ref{table: conditions for specific models and kernels}, we summarize the occurrence of pattern formation by the Turing principle for both the non-local Fisher-KPP model and the GOS model under the kernel functions we employ within the numerical simulations, i.e., for the top-hat, parabolic and cosine kernel (which have a sign-changing Fourier transform) and the triangular kernel (which has a positive Fourier transform).

\subsection{Simulation results}

The numerical method used for the model simulations, along with the specific model, kernel function and parameter values corresponding to each figure in this section, is provided in Appendix \ref{appendix: numerics}.

\subsubsection{Near the onset of Turing patterns}

\begin{figure}[b!]
    \centering
    \begin{minipage}[b]{0.49\textwidth}
        \centering
        \includegraphics[width=\textwidth]{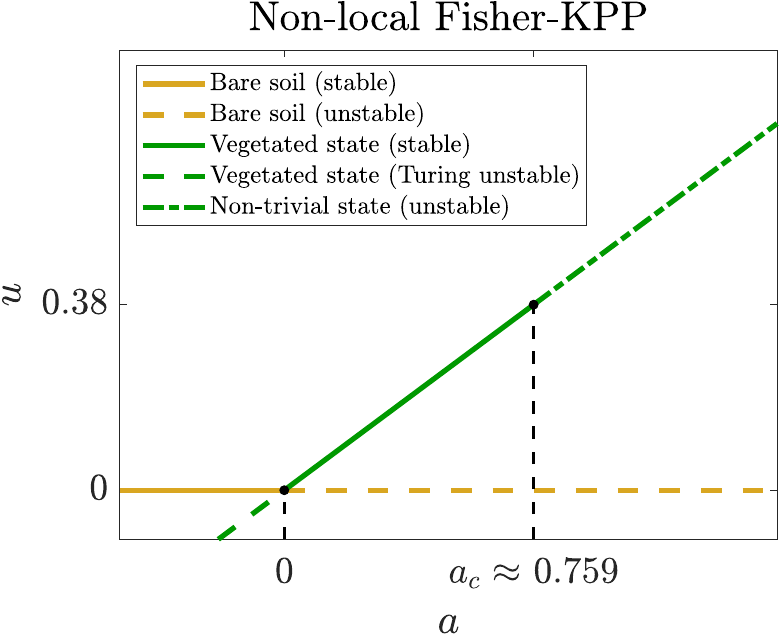}
    \end{minipage}%
    \hspace{0.15cm}
    \begin{minipage}[b]{0.49\textwidth}
        \centering
        \includegraphics[width=\textwidth]{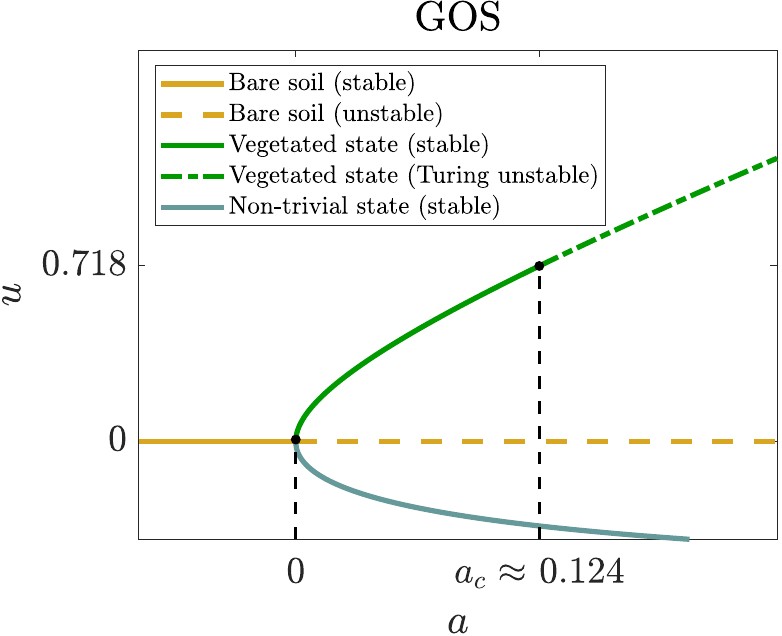}
    \end{minipage}%
\caption{Bifurcation diagrams for the non-local Fisher-KPP model (left) and the GOS model (right), both with a top-hat kernel. The bifurcation parameter is the local growth rate $a$. At $a=0$, a positive and stable vegetated steady state emerges and at $a=a_c$ this equilibrium becomes Turing unstable.}
\label{figure: bifurcation diagrams}
\end{figure}

As a first step, we compute bifurcation diagrams to visualize the stability changes of the homogeneous equilibria with varying parameters. We pick the local growth rate $a$ as the bifurcation parameter, as this choice enables a direct comparison between the two models, which share the same local growth function. The other shared parameters, the diffusion rate $D$ and the interaction range $\ell$, do not influence the biomass density of the uniform steady states and are therefore less suitable for this purpose. The resulting bifurcation diagrams for both models with a top-hat competition kernel are shown in Figure \ref{figure: bifurcation diagrams}.

The bifurcating structures of the two models exhibit qualitative similarities. For negative values of $a$ (when mortality exceeds local growth), the bare soil state $u=0$ is stable. As $a$ increases past zero, a bifurcation occurs where the bare soil state loses stability. In the non-local Fisher-KPP model, this bifurcation is transcritical and the equilibrium ($\bar{u} = \frac{a}{b}$) becomes positive and stable, representing a uniformly vegetated state. This branch increases linearly with $a$.

In the GOS model, two non-trivial, stable equilibria emerge in a pitchfork bifurcation, of which only one is positive, representing a vegetated state, namely 
\begin{align}
    \bar{u} = \frac{ac}{2b}\left(1+\sqrt{1+\frac{4b}{ac^2}}\right).
\end{align}
This homogeneous steady state curves downward as $a$ increases.

In both models, the uniformly vegetated state remains stable up to the Turing bifurcation point $a=a_c$. Beyond this threshold, the equilibrium becomes spatially unstable while remaining temporally stable for all larger values of $a$, i.e., the vegetated state is Turing unstable for all $a>a_c$.

Consistent with the theoretical result in Table \ref{table: conditions for specific models and kernels}, the non-local Fisher-KPP model with a triangular kernel does not allow for a Turing instability; instead, the vegetated equilibrium remains stable for all $a>0$. Apart from this distinction, the specific choice of the kernel and the underlying instability mechanism do not qualitatively alter the overall structure of the bifurcation diagrams. 

\begin{figure}[b!]
    \centering
    \begin{minipage}[b]{0.49\textwidth}
        \centering
        \includegraphics[width=\textwidth]{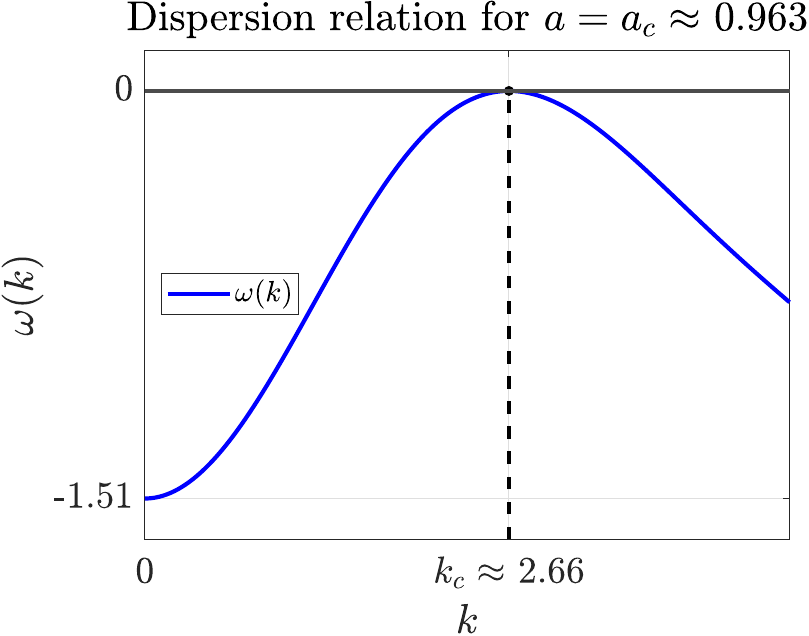}
    \end{minipage}%
    \hspace{0.15cm}
    \begin{minipage}[b]{0.49\textwidth}
        \centering
        \includegraphics[width=\textwidth]{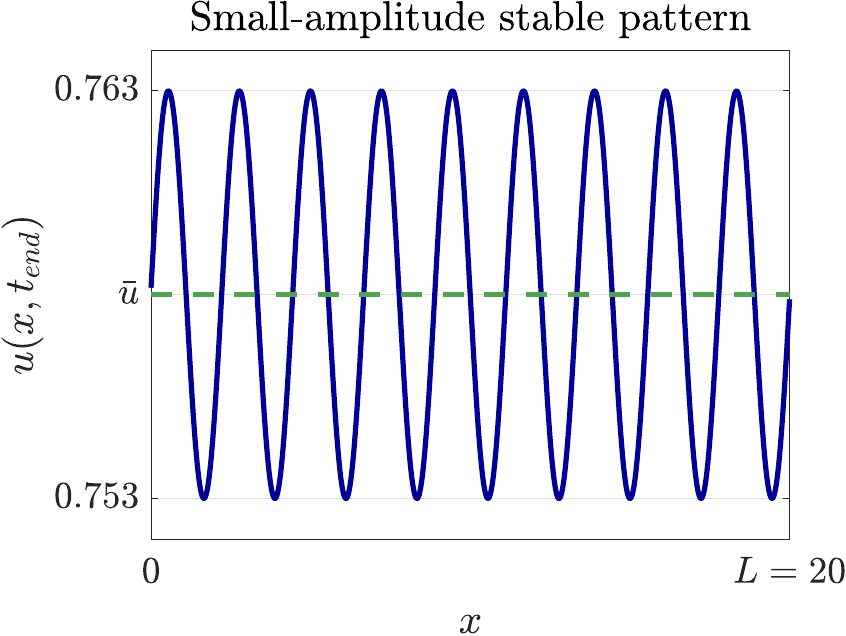}
    \end{minipage}%
\caption{Dispersion relation at $a = a_c$ (left) and numerical simulation near the onset of Turing patterns ($a \gtrsim a_c$) (right) for the GOS model with a triangular kernel. Small-amplitude stable patterns arise just beyond the Turing bifurcation point indicating a supercritical Turing bifurcation.}
\label{figure: small amplitude stable pattern}
\end{figure}

Quantitatively, several observations can be made. Under the growth-outpacing-susceptibility mechanism, kernels that decay rapidly to near zero induce a Turing bifurcation at a smaller value of $a_c$. Under the enhanced-competition-in-between-vegetated-patches mechanism, both the location and the magnitude of the minimum of $\hat{\phi}(k)$ influence $a_c$: minima that occur at smaller wavenumbers and attain larger negative values lead to smaller values of $a_c$. Consequently, the top-hat kernel triggers the earliest Turing bifurcation among the kernels considered for constant $\ell$, as can be inferred from the Fourier transforms shown in the bottom row of Figure \ref{figure: representative kernels}. Additional quantitative differences arise between the two models due to their distinct governing equations.

In the left panel of Figure \ref{figure: small amplitude stable pattern}, we display the dispersion relation evaluated at the Turing bifurcation point corresponding to the GOS model with a triangular kernel. We complement this plot in the right panel by a numerical simulation with $a$ slightly above $a_c$, illustrating the loss of stability of the uniformly vegetated state. From this, we infer that the Turing bifurcation is supercritical, as a small-amplitude stable pattern emerges immediately past the bifurcation point (see also the discussion in Section \ref{subsubsection: exclusion areas}). 

In our numerical experiments, the nature of the Turing bifurcation is consistently supercritical for both models, across different kernels and parameter settings. The resulting patterns are nearly sinusoidal and remain close to the homogeneous equilibrium $\bar{u}$. No qualitative differences are observed between the distinct mechanisms generating the Turing bifurcation.

\subsubsection{Far from homogeneous equilibrium}

\begin{figure}[t]
    \centering
    \begin{minipage}[b]{0.32\textwidth}
        \centering
        \includegraphics[width=\textwidth]{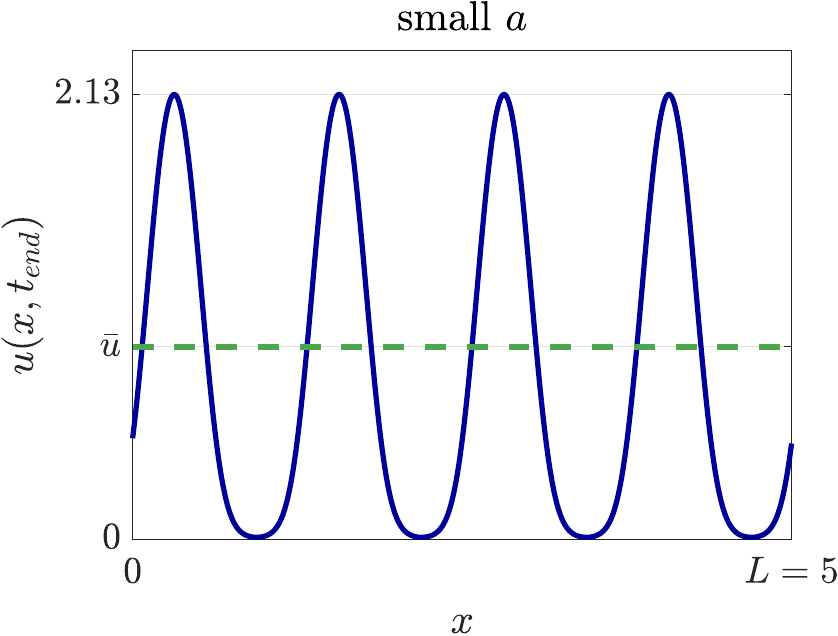}
    \end{minipage}%
    \begin{minipage}[b]{0.32\textwidth}
        \centering
        \includegraphics[width=\textwidth]{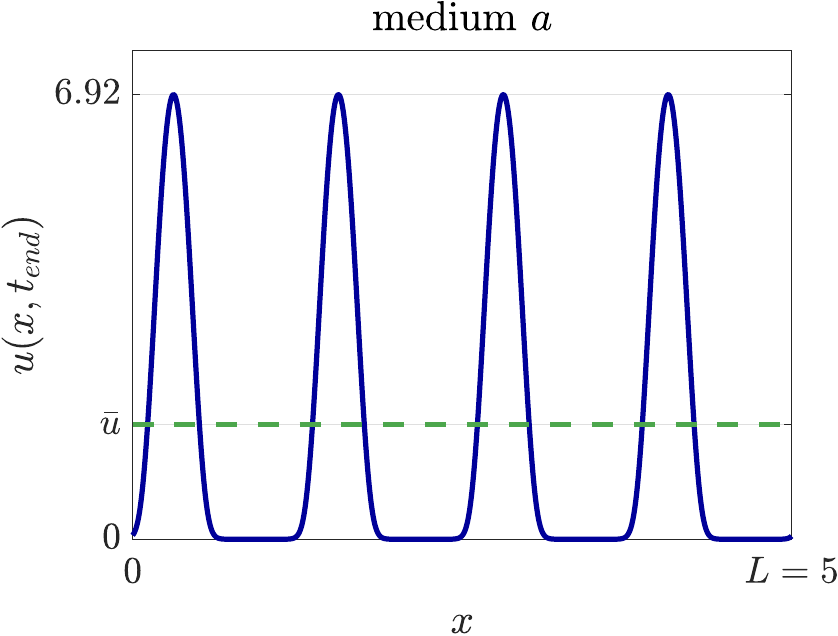}
    \end{minipage}%
    \begin{minipage}[b]{0.32\textwidth}
        \centering
        \includegraphics[width=\textwidth]{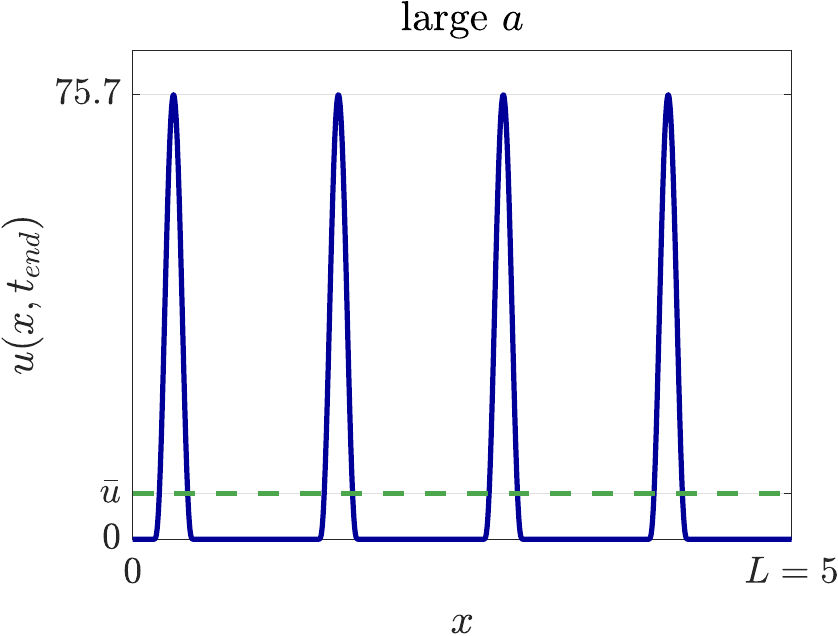}
    \end{minipage}%
\caption{Patterns far from homogeneous equilibrium ($a \gg a_c$) in the GOS model with a top-hat kernel. Increasing the local growth rate $a$ results in taller peaks and wider excursions to zero between them.}
\label{figure: infinite extent Busse balloon}
\end{figure}

As $a$ is increased further beyond $a_c$, the small-amplitude stable patterns undergo characteristic changes. Initially, the amplitude of the pulses grows while the pattern retains a nearly sinusoidal shape. This continues until the lower parts of the pulses first reach the zero axis, as shown in the left panel of Figure \ref{figure: infinite extent Busse balloon}. This indicates the emergence of bare soil regions (i.e., exclusion zones) in between the vegetated areas when the growth rate $a$ is sufficiently large. As $a$ increases even further, the pulse peaks continue to rise while the excursions to zero broaden (middle and right panels of Figure \ref{figure: infinite extent Busse balloon}). In our numerical experiments, no critical value of $a$ was observed at which the patterns collapse. This suggests pattern persistence for all $a>a_c$.

\begin{figure}[b]
    \centering
    \begin{minipage}[b]{0.32\textwidth}
        \centering
        \includegraphics[width=\textwidth]{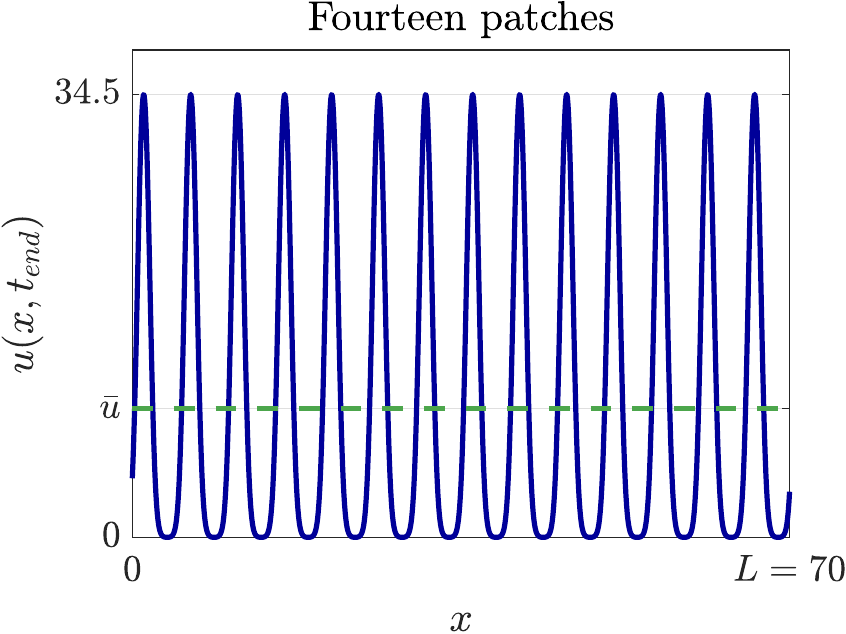}
    \end{minipage}%
    \begin{minipage}[b]{0.32\textwidth}
        \centering
        \includegraphics[width=\textwidth]{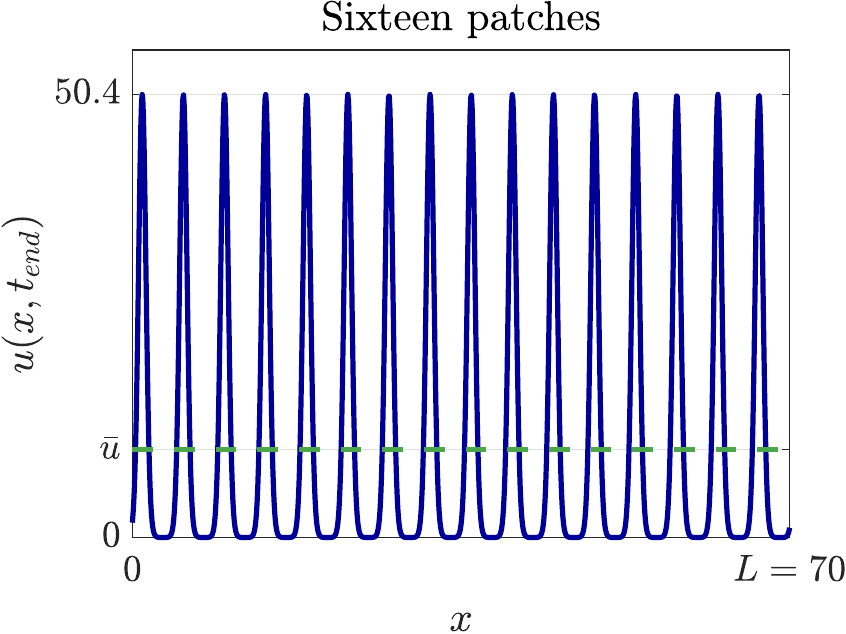}
    \end{minipage}%
    \begin{minipage}[b]{0.32\textwidth}
        \centering
        \includegraphics[width=\textwidth]{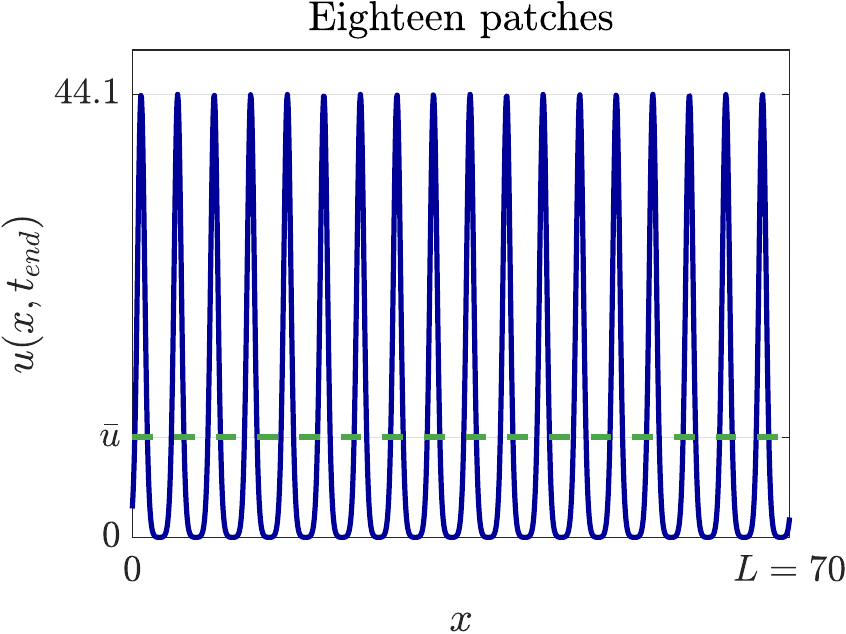}
    \end{minipage}%
\caption{Multi-stability of patterned states in the non-local Fisher-KPP model with a parabolic kernel: for a fixed parameter setting, stable patterns with different wavelengths coexist.}
\label{figure: multi-stability}
\end{figure}

Near the onset of Turing patterns, there is typically only one unstable wavenumber that fits onto a finite domain, resulting in a unique small-amplitude stable pattern for a specific parameter setting. However, for $a$ sufficiently larger than $a_c$, multiple stable patterns with different wavelengths can coexist. Figure \ref{figure: multi-stability} illustrates the multi-stability: for a fixed parameter setting, configurations with a different number of vegetated patches are stable, with the final state of the system determined entirely by the initial conditions.

Together, these observations indicate the presence of a Busse balloon with infinite extent. The Busse balloon is defined as the region in the wavenumber-parameter space (in our setting, the $(k,a)$-plane) corresponding to stable patterned states \citep{Bastiaansen2018}. Its full shape can, in principle, be determined numerically via continuation methods \citep{Uecker2012}, as has been done for reaction-diffusion systems representing dryland ecosystems \citep{Siteur2014,Vanderveken2023}. This approach can also be extended to non-local settings \citep{Rankin2014}, but such an analysis lies beyond the scope of the present work.

Our extensive numerical exploration indicates that the existence of the infinite Busse balloon is independent of the specific instability mechanism generating the Turing bifurcation: we recover this structure across both models, for all parameter settings and kernel choices that permit a Turing instability. This suggests that the distinct competition-induced instability mechanisms can give rise to vegetation patterns with qualitatively similar behaviour.

\begin{figure}[t]
    \centering
    \begin{minipage}[b]{0.32\textwidth}
        \centering
        \includegraphics[width=\textwidth]{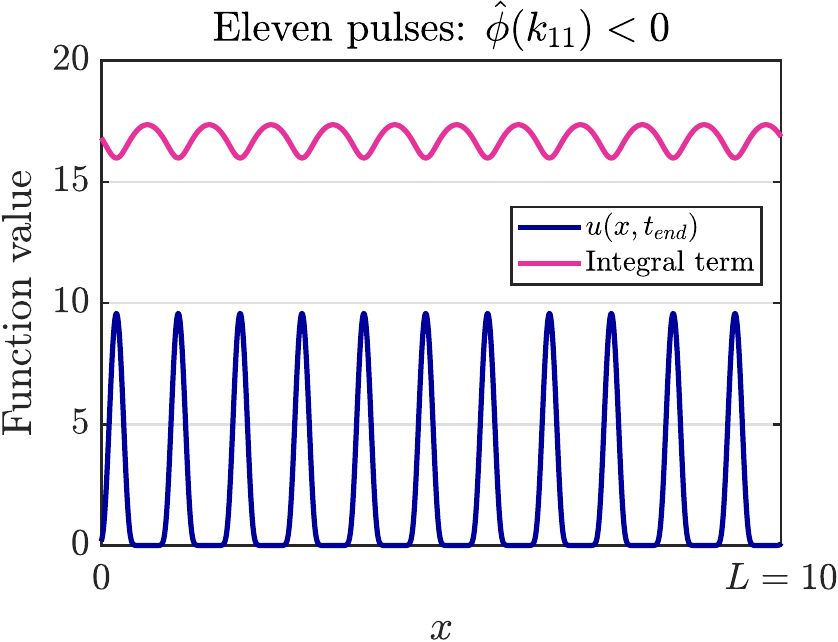}
    \end{minipage}%
    \begin{minipage}[b]{0.32\textwidth}
        \centering
        \includegraphics[width=\textwidth]{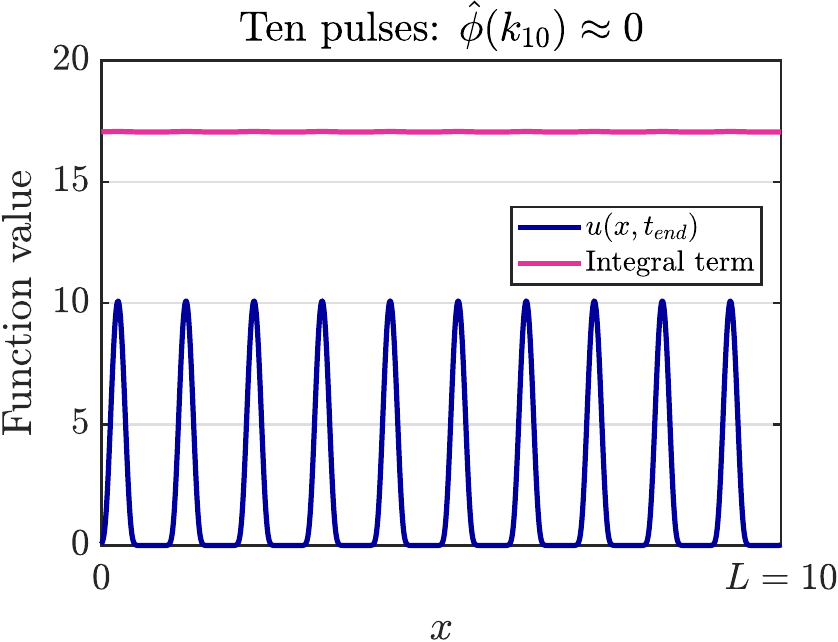}
    \end{minipage}%
    \begin{minipage}[b]{0.32\textwidth}
        \centering
        \includegraphics[width=\textwidth]{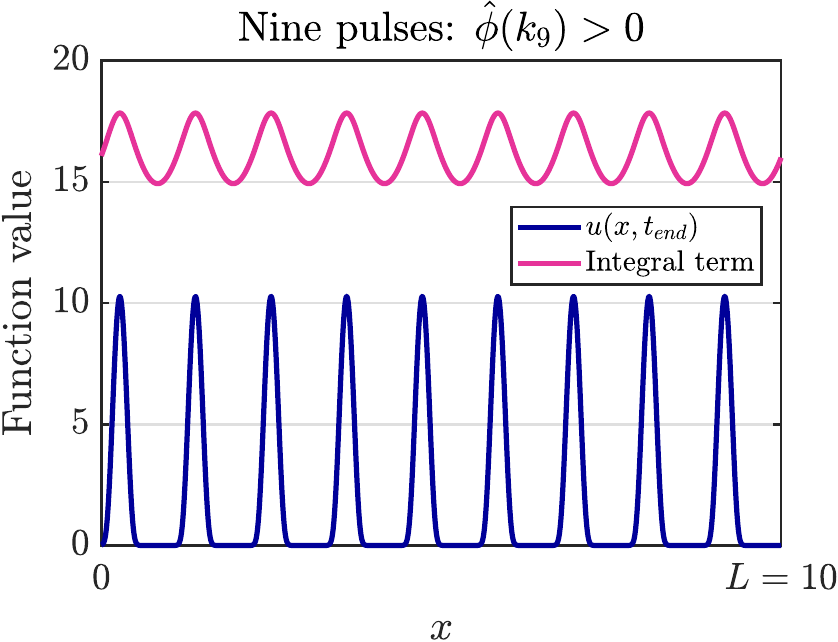}
    \end{minipage}%
\caption{The phase of the competitive impact is determined by the sign of $\hat{\phi}(k)$: for $\hat{\phi}(k) < 0$, competitive pressure is dominating in between patches (left panel); for $\hat{\phi}(k) \approx 0$, competition is nearly evenly distributed (center panel); for $\hat{\phi}(k) > 0$, competitive impact is strongest within patches (right panel). The simulations are based on the GOS model with a cosine kernel.}
\label{figure: phase and competition}
\end{figure}

However, the underlying competitive dynamics leading to those patterns is qualitatively different. In particular, the spatial phase of the competitive pressure relative to the distribution of the biomass is determined entirely by the sign of $\hat{\phi}(k)$ (Figure \ref{figure: phase and competition}). Based on a large number of numerical experiments, we conjecture the following: for any stable pattern with corresponding wavenumber $k$, whenever $\hat{\phi}(k)<0$, the competitive pressure is strongest in the low-biomass regions in between the high-biomass areas, as competition arrives from either side. Close to the bifurcation point this asymmetry is relatively weak, but as $a$ is shifted away from the onset of patterning, the contrast intensifies, leading to far-from-homogeneous-equilibrium patterns that exhibit excursions to zero, i.e., to the emergence of fully developed exclusion zones. This effect is consistent with the exclusion zone perspective of \citep{Pigolotti2007,Pigolotti2010,Martinez2013,Martinez2013A} and independent of the sign of derivative of the ratio $\frac{g(u)}{s(u)}$.

Conversely, if $\hat{\phi}(k)>0$, the competitive impact is strongest within the high-biomass areas, driven by self-limiting interactions within these regions. This contrasts with the aforementioned exclusion zone interpretation, and patterns in this regime must therefore be governed by the growth-outpacing-susceptibility mechanism. In the transitional case $\hat{\phi}(k) \approx 0$, the competitive impact is distributed roughly uniformly across space, while the corresponding pattern remains robustly stable.

Importantly, this phase distinction is not restricted to the immediate vicinity of the Turing bifurcation, but persists well beyond the onset of patterning. This indicates that the sign of $\hat{\phi}(k)$ is a key factor in shaping pattern dynamics not only near the Turing bifurcation (as indicated by our linear stability analysis), but throughout the entire parameter space.

\section{Discussion}

In this work, we introduced and analyzed a general framework for vegetation dynamics in arid and semi-arid ecosystems subject to non-local self-limiting competition. Through linear stability analysis, we identified two distinct mechanisms that can lead to the emergence of spatial patterns via the Turing principle when the rate of biomass spread is sufficiently low. The first mechanism requires the Fourier transform of the competition kernel to satisfy $\hat{\phi}(k)<0$ for some range of wavenumbers $k$, thereby allowing for strong competition in between more densely vegetated areas due to the cumulative competitive effects. The second mechanism requires growth to outpace competitive susceptibility in the vicinity of the homogeneous equilibrium. This condition aligns with the local activation component of the LALI-principle, while the long-range inhibition is governed by non-local competition. Notably, this mechanism enables pattern formation under any ecologically relevant competition kernel, without additional restrictions.

The first mechanism has been derived and explored in several previous studies \citep{Fuentes2004,Segal2013,Martinez2013A,Martinez2013,Martinez2014,Tega2022}, while the second mechanism is typically absent in the literature due to the use of simplified functional responses. Our results demonstrate that incorporating the empirically observed non-linear nature of ecological interactions can reveal an alternative pathway to pattern formation when non-local interactions are purely competitive. We therefore emphasize the importance of occasionally stepping back from prescribing specific functional forms early in the modelling process and instead considering more general frameworks, which can uncover key mechanisms that might otherwise remain hidden.

We complemented our analytical results with numerical simulations, which showed that qualitatively similar stable Turing patterns can emerge from the two instability mechanisms. The spatial phase of the competitive impact relative to the biomass distribution depends on the sign of $\hat{\phi}(k)$: positive values indicate that competitive effects are strongest within densely vegetated patches (consistent with the exclusion area perspective), while negative values shift competitive pressure towards the regions in between these patches (aligning with the growth-outpacing-susceptibility mechanism). This result holds both near the onset of vegetation patterning and in the far-from-homogeneous-equilibrium regime. It therefore provides a way to identify the governing pattern-forming mechanism by comparing the competitive pressure, for example through water availability, between vegetated patches and bare soil regions.

The supercritical character of the Turing bifurcation and the structure of the Busse balloon observed in our numerical simulations depend on the specific functional forms chosen in the benchmark models. Different choices for these functions can give rise to subcritical Turing bifurcations or finite Busse balloons. Specifically, under different modelling choices, our framework can exhibit the so-called Turing-resists-tipping mechanism \citep{Rietkerk2021,Voort2025}, in which stable patterned states persist under parameter changes that would drive the corresponding mean-field system towards collapse. In these scenarios, spatial self-organization acts as a resilience mechanism. Therefore, the framework is not limited to the behaviour presented in this study but also provides a basis for examining ecosystem resilience, as has been explored more extensively in reaction-diffusion systems \citep{Siteur2014,Bastiaansen2018}.

Furthermore, future work could extend our near-onset analysis by deriving the amplitude equations corresponding to the Turing bifurcation. Specifically, by computing the Landau coefficient one can distinguish between supercritical bifurcations (yielding small-amplitude stable patterns, as observed in this study) and subcritical bifurcations (inducing transient patterning, potentially leading to localized structures or tipping) analytically, depending on the specific parameter setting \citep{vanderstelt2013,Paulau2014,Doelman2019,Krause2024,Voort2025}. Although deriving amplitude equations is technically challenging, especially for non-local systems, recent work provides some promising progress \citep{Garlaschi2021,Pal2022}.

In this study, we relied on numerical simulations to explore the behaviour of patterned solutions far from onset. For the localized patterns observed, application of geometric singular perturbation theory could yield analytical insights into the underlying dynamics and help to prove some of our claims rigorously. However, this method has been primarily applied to local reaction-diffusion(-advection) systems and would require extension to be applicable to non-local models. An alternative approach would be to construct a (near-)equivalent reaction-diffusion formulation of our model, following previous efforts that have established general connections between local and non-local frameworks \citep{Ninomiya2017,PintoRamos2023,PintoRamos2025}. Moreover, developing an analytical understanding of the intermediate region connecting onset to far-from-onset dynamics would help to understand the evolution of transitioning patterns, as has already been achieved in a local setting \citep{Morgan2000,Brown2023}.

The general framework presented in this study can be extended in several ways to increase ecological realism. The kernel function $\phi(x)$ effectively represents the root system of a plant, which is dynamic over its lifetime and depends both on plant size and on the local density of self and non-self roots \citep{Cabal2020}. In the current formulation, its interaction range $\ell$ is fixed, but this distance should vary with the local biomass density. Introducing an allometric scaling for $\ell$, as done in \citep{Tlidi2024}, would enable us to investigate how a dynamic interaction range influences pattern onset and dynamics, and whether it gives rise to additional mechanisms that generate spatial instabilities. These questions are the subject of ongoing research. Alternatively, the symmetry property imposed on $\phi(x)$ could be relaxed, allowing the description of scenarios in which anisotropic environmental conditions lead to non-reciprocal plant interactions \citep{PintoRamos2025}.

While we focused on a one-dimensional spatial domain to identify the key ingredients underlying each pattern-forming mechanism, extending the framework to two spatial dimensions is essential to bridge the gap between theoretical modelling and applications to real ecosystems. In two dimensions, a rich variety of pattern configurations exists, including spots, labyrinths, stripes and gaps, as reported for non-local vegetation models in \citep{Martinez2014}. In contrast, the non-local Fisher-KPP model subjected to periodic boundary conditions tends to only generate vegetation spots \citep{Silvano2025}. Thus, although the kernel structure is key to enabling pattern formation, the specific nonlinearities determine which configurations are attainable. Future studies may employ our general framework to analyze how particular choices of the functional forms influence pattern selection in two spatial dimensions.

Another natural extension is to consider a multi-component system that distinguishes between different types of vegetation \citep{Andreguetto2021}. For instance, a model with two woody plant communities, such as trees and shrubs, each with distinct ranges of lateral spread, would allow us to investigate how non-local competitive interactions both within and between these groups affect the emergence and dynamics of vegetation patterns. Adding a herbaceous component would further enable the study of coexistence between woody plants and grasses, which differ more strongly in their functional traits. In particular, grasses exert competition over much shorter spatial scales and may even benefit from facilitation by trees through reduced water evaporation due to shading \citep{Scholes1997,Schenk2002,Moustakas2013}. Taken together, these extensions would provide a more realistic representation of a savanna ecosystem.

To summarize, this work generalizes and extends previous modelling studies on non-local self-limiting competition and its role in shaping the formation and dynamics of vegetation patterns. We hope that the framework introduced here will serve as a basis for future efforts to advance our understanding of heterogeneous ecosystems and their governing dynamics.

\section{Acknowledgments}
This research was supported by the Dutch Research Council (NWO) through the project ``Resilience in Complex Systems through Adaptive Spatial Pattern Formation'' (project number OCENW.M20.169), with contributions from J.V. and A.D. Additional funding for A.D. was provided by the European Research Council through the ERC-Synergy project RESILIENCE (proposal number 101071417). This work was partially funded by the Center of Advanced Systems Understanding (CASUS), which is financed by Germany’s Federal Ministry of Research, Technology and Space (BMFTR) and by the Saxon Ministry for Science, Culture and Tourism (SMWK) with tax funds on the basis of the budget approved by the Saxon State Parliament. RMG was also supported by the São Paulo Research Foundation (FAPESP) through grant ICTP-SAIFR 2021/14335-0.

\begin{appendices}

\section{Kernel functions}
\label{appendix: kernel functions}

\subsection{Top-hat kernel}

\underline{Kernel set-up}
\vspace{2mm}\\
We start with a kernel ansatz of the form:
\begin{align}
\phi(x) = 
\begin{cases}
\displaystyle A & \text{if } |x| < \ell, \\
0 & \text{if } |x| > \ell,
\end{cases}
\end{align}
with $A,\ell >0$. The normalization property gives:
\begin{align}
    1 = 2A \int^\ell_0 1\, dx = 2A\ell,
\end{align}
which yields $A = \frac{1}{2\ell}$. Thus, the top-hat kernel is given by:
\begin{align}
\phi(x) = 
\begin{cases}
\displaystyle\frac{1}{2\ell} & \text{if } |x| < \ell, \\
0 & \text{if } |x| > \ell,
\end{cases}
\end{align}
with $\ell>0$ the interaction range.
\vspace{20mm}\\
\underline{Fourier transform}
\vspace{2mm}\\
We compute the Fourier transform (\ref{definition: FT of phi}) for $k \neq 0$:
\begin{align}
    \hat{\phi}(k) &= \frac{1}{\ell}\int^\ell_{0} \cos(kx)\, dx = \frac{\sin(k\ell)}{k\ell}.
\end{align}
Combining this with $\hat{\phi}(0) = 1$, we conclude the Fourier transform of the top-hat kernel to be given by:
\begin{align}
    \hat{\phi}(k) = \sinc(k\ell).
\end{align}

\subsection{Parabolic kernel}
\underline{Kernel set-up}
\vspace{2mm}\\
We start with a kernel ansatz of the form:
\begin{align}
\phi(x) = 
\begin{cases}
\displaystyle A+Bx+Cx^2 & \text{if } |x| \leq \ell, \\
0 & \text{if } |x| > \ell,
\end{cases}
\end{align}
with $A,B,C \in \mathbb{R}$ and $\ell >0$. Continuity requires $\phi(x)$ to equal zero at $x = \pm \ell$. This leads to the system of equations:
\begin{align}
    0 & = A +B\ell +C\ell^2,\\
    0 &= A- B\ell +C\ell^2.
\end{align}
Solving this system yields $B=0$ and $C = -\frac{A}{\ell^2}$. The normalization property gives:
\begin{align}
    1 = 2A \int^\ell_0 1- \frac{x^2}{\ell^2}\, dx = 2A\left(\ell-\frac{\ell}{3}\right) = \frac{4A\ell}{3},
\end{align}
which yields $A = \frac{3}{4\ell}$. Thus, the parabolic kernel is given by:
\begin{align}
\phi(x) = 
\begin{cases}
\displaystyle \frac{3}{4\ell}\left(1-\left(\frac{x}{\ell}\right)^2\right) & \text{if } |x| \leq \ell, \\
0 & \text{if } |x| > \ell,
\end{cases}
\end{align}
with $\ell>0$ the interaction range.
\vspace{2mm}\\
\underline{Fourier transform}
\vspace{2mm}\\
We compute the Fourier transform (\ref{definition: FT of phi}) for $k \neq 0$:
\begin{align}
    \hat{\phi}(k) &= \frac{3}{2\ell} \int^\ell_{0} \left(1-\left(\frac{x}{\ell}\right)^2\right) \cos(kx)\,dx\\
    &= \frac{3}{2\ell} \left(\frac{\sin(k\ell)}{k}-\frac{1}{\ell^2}\left(\left[\frac{x^2\sin(kx)}{k}\right]^{x=\ell}_{x=0}-\frac{2}{k}\int^\ell_{0}x\sin(kx)\,dx\right)\right)\\
    &= \frac{3}{k\ell^3} \left(\left[-\frac{x\cos(kx)}{k}\right]^{x=\ell}_{x=0}+\frac{1}{k}\int^\ell_{0}\cos(kx)\,dx\right)\\
    &= \frac{3}{k^2\ell^2} \left(-\cos(k\ell)+\frac{\sin(k\ell)}{k\ell}\right).
\end{align}
Combining this with $\hat{\phi}(0) = 1$, we conclude the Fourier transform of the parabolic kernel to be given by:
\begin{align}
    \hat{\phi}(k) = 
\begin{cases}
\displaystyle \frac{3}{(k\ell)^2}(\sinc(k\ell)-\cos(k\ell)) & \text{if } k \neq 0,\\
1 & \text{if } k = 0,\\
\end{cases}
\end{align}

\subsection{Cosine kernel}
\underline{Kernel set-up}
\vspace{2mm}\\
We start with a kernel ansatz of the form:
\begin{align}
\phi(x) = 
\begin{cases}
\displaystyle A\cos(Bx+C)+D & \text{if } |x| \leq \ell, \\
0 & \text{if } |x| > \ell,
\end{cases}
\end{align}
with $A,B,C,D \in \mathbb{R}$ and $\ell > 0$. For the kernel to be symmetric, we must have $C=0 \mod 2\pi$. At the boundary points $x = \pm \ell$, the cosine should be at its minimum and for continuity $\phi(x)$ should equal zero. This yields $B=\frac{\pi}{\ell}$ and $D=A$. The normalization property gives:
\begin{align}
    1 = 2A \int^\ell_{0} \cos\left(\frac{\pi}{\ell}x\right)+1\, dx = 2A \left[\frac{\ell}{\pi}\sin\left(\frac{\pi}{\ell}x\right)+x\right]^{x=\ell}_{x=0} = 2A\ell,
\end{align}
which yields $A = \frac{1}{2\ell}$. Thus, the cosine kernel is given by:
\begin{align}
\phi(x) = 
\begin{cases}
\displaystyle\frac{1}{2\ell}\left(\cos\left(\frac{\pi}{\ell}x\right)+1\right) & \text{if } |x| \leq \ell, \\
0 & \text{if } |x| > \ell.
\end{cases}
\end{align}
with $\ell>0$ the interaction range.
\vspace{2mm}\\
\underline{Fourier transform}
\vspace{2mm}\\
We compute the Fourier transform (\ref{definition: FT of phi}) for $k \neq 0$:
\begin{align}
    \hat{\phi}(k) &= \frac{1}{\ell} \int^\ell_{0} \left(\cos\left(\frac{\pi}{\ell}x\right)+1\right) \cos(kx)\,dx\\
    &= \frac{1}{\ell} \int^\ell_{0} \cos\left(\frac{\pi}{\ell}x\right) \cos(kx)\,dx + \frac{\sin(k\ell)}{k\ell}
\end{align}
For $k = \pm\frac{\pi}{\ell}$, we find:
\begin{align}
    \hat{\phi}\left(\pm\frac{\pi}{\ell}\right) &= \frac{1}{\ell} \int^\ell_{0} \cos^2\left(\frac{\pi}{\ell}x\right)\,dx = \frac{1}{2\ell} \int^\ell_{0} \cos\left(\frac{2\pi}{\ell}x\right)+1\, dx = \frac{1}{2}.
\end{align}
For $k \neq \pm\frac{\pi}{\ell}$, we find:
\begin{align}
    \hat{\phi}(k) &= \frac{1}{\ell} \left[\frac{\frac{\pi}{\ell}\sin\left(\frac{\pi}{\ell}x\right)\cos(kx)-k\cos\left(\frac{\pi}{\ell}x\right)\sin(kx)}{\left(\frac{\pi}{\ell}\right)^2-k^2}\right]^{x=\ell}_{x=0}  + \frac{\sin(k\ell)}{k\ell}\\
    &= \frac{1}{\ell} \cdot \frac{k\sin(k\ell)}{\left(\frac{\pi}{\ell}\right)^2-k^2}  + \frac{\sin(k\ell)}{k\ell}\\
    &= \frac{\pi^2}{\pi^2-(k\ell)^2}\frac{\sin(k\ell)}{k\ell}
\end{align}
Combining this with $\hat{\phi}(0) = 1$, we conclude the Fourier transform of the cosine kernel to be given by:
\begin{align}
    \hat{\phi}(k) = 
\begin{cases}
\displaystyle \frac{\pi^2}{\pi^2-(k\ell)^2} \sinc(k\ell) & \text{if } |k| \neq \frac{\pi}{\ell},\\
\frac{1}{2} & \text{if } |k| = \frac{\pi}{\ell},\\
\end{cases}
\end{align}

\subsection{triangular kernel}
\underline{Kernel set-up}
\vspace{2mm}\\
We start with a kernel ansatz of the form:
\begin{align}
\phi(x) = 
\begin{cases}
\displaystyle A-B\left|x\right| & \text{if } |x| \leq \ell, \\
0 & \text{if } |x| > \ell.
\end{cases}
\end{align}
with $A,B,\ell >0$. Continuity requires $\phi(x)$ to equal zero at $x = \pm \ell$. This yields $B= \frac{A}{\ell}$. The normalization property gives:
\begin{align}
    1 = 2A\int^\ell_{0} 1-\frac{x}{\ell}\, dx = 2A \left(\ell - \frac{\ell}{2}\right) = A \ell,
\end{align}
which yields $A = \frac{1}{\ell}$. Thus, the triangular kernel is given by:
\begin{align}
\phi(x) = 
\begin{cases}
\displaystyle\frac{1}{\ell}\left(1-\frac{\left|x\right|}{\ell}\right) & \text{if } |x| \leq \ell, \\
0 & \text{if } |x| > \ell.
\end{cases}
\end{align}
with $\ell>0$ the interaction range.
\vspace{2mm}\\
\underline{Fourier transform}
\vspace{2mm}\\
We compute the Fourier transform (\ref{definition: FT of phi}) for $k \neq 0$:
\begin{align}
    \hat{\phi}(k) &= \frac{2}{\ell} \int^\ell_{0} \left(1-\frac{x}{\ell}\right) \cos(kx)\,dx\\
    &= \frac{2}{\ell} \left(\frac{\sin(k\ell)}{k}-\frac{1}{\ell}\left(\left[\frac{x\sin(kx)}{k}\right]^{x=\ell}_{x=0}-\frac{1}{k}\int^\ell_{0}\sin(kx)\,dx\right)\right)\\
    &= \frac{2}{k^2\ell^2}(1-\cos(k\ell))\\ &= \frac{4}{k^2\ell^2}\sinc^2\left(\frac{k\ell}{2}\right).
\end{align}
Combining this with $\hat{\phi}(0) = 1$, we conclude the Fourier transform of the triangular kernel to be given by:
\begin{align}
    \hat{\phi}(k) = \sinc^2\left(\frac{k\ell}{2}\right),
\end{align}
which is non-negative for all $k$.

\section{Numerical setup}
\label{appendix: numerics}

All figures were generated using Matlab. For the simulations, the one-dimensional spatial domain was discretized using the Method of Lines. The integral term in the non-local contribution was approximated by a finite sum, and periodic boundary conditions were imposed to minimize boundary effects. Time integration was performed using Matlab's ode15s solver. The initial conditions consisted either of a slightly perturbed homogeneous equilibrium state or of a sinusoidal profile chosen to enforce a prescribed number of pulses.

Table \ref{table: numerical setup} lists the models, kernel functions and parameter values used to create the figures in Section \ref{section: numerics}.

\begin{table}[h!]
\centering
\renewcommand{\arraystretch}{1.3}
\begin{tabular}{lcccccccc}
\toprule
 & Model & Kernel & $a$ & $b$ & $c$ & $D$ & $\ell$ & $L$ \\
\midrule
Figure 6 (left) & FKPP & top-hat & $-$ & $2$ & $-$ & $0.009$ & $1$ & $-$ \\
Figure 6 (right) & GOS & top-hat & $-$ & $2$ & $3$ & $0.009$ & $1$ & $-$ \\
Figure 7 & GOS & triangular & $0.9804$ & $3$ & $1$ & $0.05$ & $2$ & $20$ \\
Figure 8 & GOS & top-hat & $0.45$/$1$/$5$ & $2$ & $3$ & $0.009$ & $1$ & $5$ \\
Figure 9 & FKPP & parabolic & $4$ & $0.4$ & $-$ & $0.02$ & $4$ & $70$ \\
Figure 10 & GOS & cosine & $2$ & $2$ & $3$ & $0.009$ & $1$ & $10$ \\
\bottomrule
\end{tabular}
\caption{Numerical setup for all figures presented in Section \ref{section: numerics}. A dash indicates that a parameter is not applicable to the respective simulation.}
\label{table: numerical setup}
\end{table}
\end{appendices}

\bibliographystyle{unsrtnat} 
\bibliography{ref}

\end{document}